\documentclass[12pt]{article}
\usepackage{epsfig, amsmath, amssymb}
\usepackage{graphicx,psfrag}
\usepackage{xcolor} 
\renewcommand{\baselinestretch}{1.23}

\usepackage{microtype}
\usepackage{kantlipsum}

\newcommand{\nn}{\nonumber}
\newcommand{\be}{\begin{equation}}
\newcommand{\ee}{\end{equation}}
\newcommand{\ben}{\begin{equation}}
\newcommand{\een}{\end{equation}}
\newcommand{\bea}{\begin{eqnarray}}
\newcommand{\eea}{\end{eqnarray}}
\newcommand{\bA}{\begin{array}}
\newcommand{\eA}{\end{array}}
\newcommand{\bc}{\begin{center}}
\newcommand{\ec}{\end{center}}
\newcommand{\al}{\alpha}

\newcommand{\ra}{\rightarrow}
\newcommand{\del}{\partial}

\newcommand{\ie}{{\it i.e.}}
\newcommand{\eg}{{\it e.g.}}

\newcommand{\lan}{\langle}
\newcommand{\ran}{\rangle}

\begin{document}


\begin{titlepage}

\bc

\hfill 

\vspace{25mm}


{\Huge Small Schwarzschild de Sitter black holes, \\ [2mm]
quantum extremal surfaces and islands}
\vspace{16mm}

{\large  Kaberi Goswami,\ \ K.~Narayan} \\
\vspace{3mm}
{\small \it Chennai Mathematical Institute, \\}
{\small \it SIPCOT IT Park, Siruseri 603103, India.\\}

\ec
\vspace{30mm}

\begin{abstract}
  We study 4-dimensional Schwarzschild de Sitter black holes in the
  regime where the black hole mass is small compared with the de
  Sitter scale. Then the de Sitter temperature is very low compared
  with that of the black hole and we study the black hole,
  approximating the ambient de Sitter space as a frozen classical
  background. We consider distant observers in the static diamond, far
  from the black hole but within the cosmological horizon. Using
  2-dimensional tools, we find that the entanglement entropy of
  radiation exhibits linear growth in time, indicative of the
  information paradox for the black hole. Self-consistently including
  an appropriate island emerging at late times near the black hole
  horizon leads to a reasonable Page curve. There are close parallels
  with flat space Schwarzschild black holes in the regime we consider.
\end{abstract}

\end{titlepage}

{\tiny 
\begin{tableofcontents}
\end{tableofcontents}
}

\vspace{2mm}

\section{Introduction}

The black hole information paradox \cite{Hawking:1976ra} can be
regarded as the tension between the apparent unbounded growth of
entanglement entropy of thermal Hawking radiation
\cite{Hawking:1975vcx} outside the black hole and the expectation from
quantum mechanics that entanglement entropy must become small at late
times if purity of the original matter state is to be recovered,
\ie\ it must follow the Page curve \cite{Page:1993wv,Page:2013dx}.
See \eg\ \cite{Mathur:2009hf,Almheiri:2012rt} for discussions of
various aspects of the information paradox.  Recent exciting
discoveries unravelled via the study of entanglement and quantum
extremal surfaces
\cite{Penington:2019npb,Almheiri:2019psf,Almheiri:2019hni,Penington:2019kki,Almheiri:2019qdq}
have found that including nontrivial ``island'' contributions does in
fact do this job. Quantum extremal surfaces are extrema of the
generalized gravitational entropy
\cite{Faulkner:2013ana,Engelhardt:2014gca} obtained from the classical
area of the entangling RT/HRT surface
\cite{Ryu:2006bv}-\cite{Rangamani:2016dms} after incorporating the
bulk entanglement entropy of matter, with explicit calculation
possible in effective 2-dimensional models where 2-dim CFT techniques
enable detailed analysis of the bulk entanglement entropy. The island,
arising as a nontrivial solution to extremization (near the black hole
horizon, and only at late times), reflects new replica wormhole
saddles \cite{Penington:2019kki,Almheiri:2019qdq} and serves to purify
the early Hawking radiation thereby leading to the entanglement
entropy decreasing.  There is a large body of literature on various
aspects of these issues, reviewed in
\eg\ \cite{Almheiri:2020cfm,Raju:2020smc,Chen:2021lnq}: see
\eg\ \cite{Almheiri:2019yqk}-\cite{Tian:2022pso} for a partial list of
investigations on black holes in this regard. Scrutinies of the island
formulation and alternative perspectives appear in
\eg\ \cite{Laddha:2020kvp,Geng:2021hlu,Bena:2022rna}.

\begin{figure}[h] 
\bc\includegraphics[width=30pc]{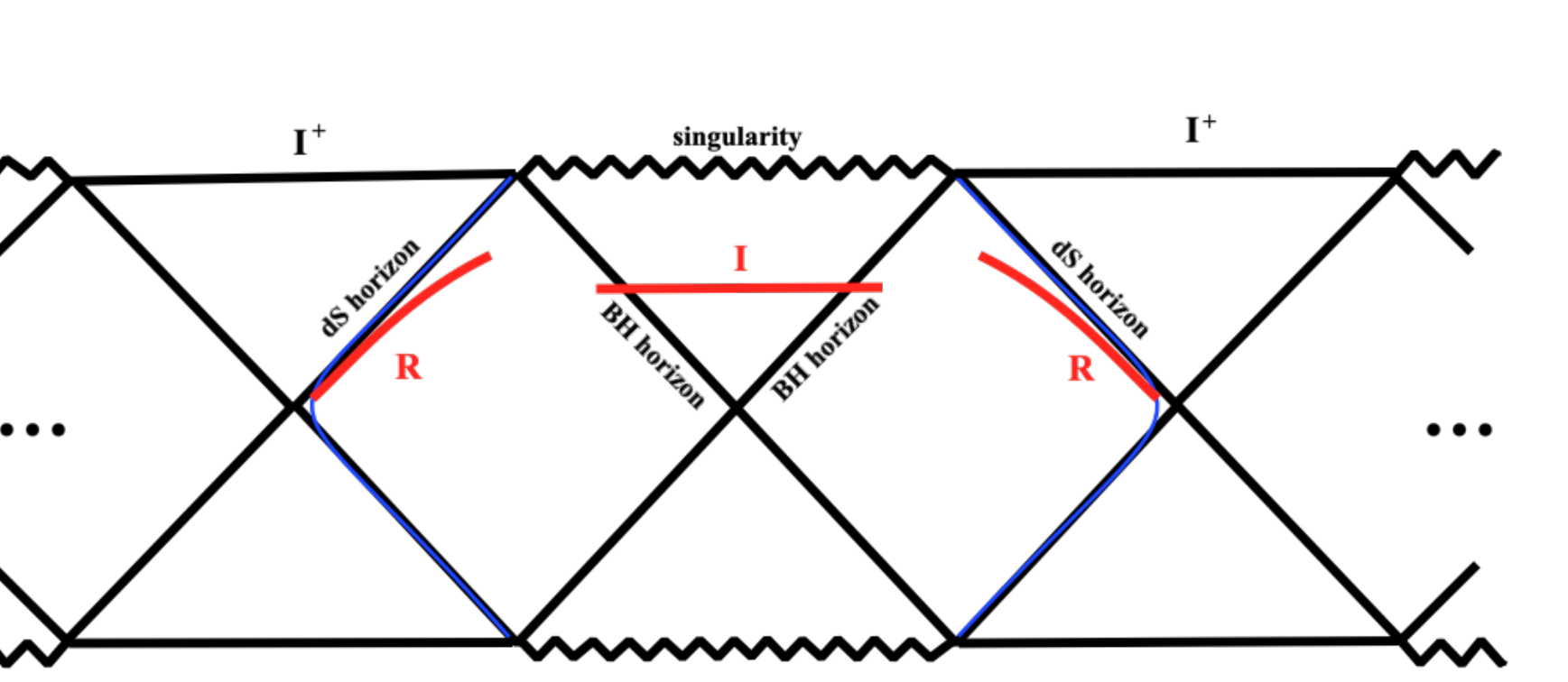}\ec \vspace{-5mm}
\caption{{ \label{figSdSqesI}
    \footnotesize{The Penrose diagram of a Schwarzschild de Sitter
      black hole, with focus on one black hole ``patch'' bounded by
      cosmological horizons on both sides.
      Depicted are the radiation regions $R_- \equiv [r_{D-},b_-]$ and
      $R_+ \equiv [b_+,r_{D+}]$ and the late time island
      $I \equiv [a_-,a_+]$.   
}}}
\end{figure}

In this paper, we study ``small'' Schwarzschild de Sitter black holes,
\ie\ the regime where the black hole mass $m$ is small compared with
the de Sitter scale $l$, but large enough that a quasi-static
approximation to the geometry is valid. This translates to saying that
the de Sitter temperature is very low compared with that of the black
hole.  In this regime, we approximate the ambient de Sitter space as a
frozen classical background and study the two-sided (eternal) black
hole. We can imagine that the black hole has formed from initial
matter in a pure state: strictly speaking this can only be an
approximation to the bulk CFT at the thermal state at the de Sitter
temperature, but it is a reasonable approximation if the de Sitter
temperature is very low. We focus on one black hole coordinate patch
in the Penrose diagram, Figure~\ref{figSdSpD} (which roughly comprises
a line of alternating Schwarzschild and de Sitter patches), and
consider observers in the static diamond patches far outside the black
hole but within the cosmological horizons which bound the black hole
patch, Figure~\ref{figSdSqesI}.  Then the entanglement entropy of the
radiation exhibits an unbounded linear growth in time, which is
inconsistent at late times with the entropy of the black hole, and
indicative of the information paradox for the black hole. Using the
island rule in the extremization of the generalized entropy shows an
island emerging at late times a little outside the black hole horizon
semiclassically: this then shows finiteness of the entanglement
entropy of radiation recovering the expectations on the Page curve.
In some essential sense, our analysis (which is purely bulk, with no
holography per se) closely mirrors island studies of flat space
Schwarzschild black holes in the literature, with the ambient de
Sitter space entering only through more complicated coordinates.

In sec.~\ref{sec:SdSrev}, we review 4-dim Schwarzschild de Sitter
black holes as required for our purposes, and in
sec.~\ref{sec:SdSbhEE}, we describe our setup for the generalized
entropy via 2-dim techniques.  Sec.~\ref{sec:noIsl} discusses the
entanglement entropy in the absence of the island (with details in
App.~\ref{App:noIsland}), while sec.~\ref{sec:EEisland} discusses the
island calculation (details in App.~\ref{App:Island}; see also
App.~\ref{App:early time entropy } for early times).  Finally
sec.~\ref{sec:Disc} contains a Discussion of our approximations and
open questions.

\section{Small Schwarzschild de Sitter black holes\ $\ra$\
  2-dim}\label{sec:SdSrev}

The Schwarzschild de Sitter black hole spacetime in $3+1$-dimensions
has the metric
\be\label{SdSst}
ds^2= -f(r)dt^2+\frac{dr^2}{f(r)}+r^2d\Omega_2^2\ ,  \qquad  
f(r)=1-\frac{2m}{r}-\frac{r^2}{l^2}\ .
\ee
This is a Schwarzschild black hole in de Sitter space
\cite{Gibbons:1977mu} with an ``outer'' cosmological (de Sitter)
horizon and an ``inner'' Schwarzschild horizon. The surface gravity at both
horizons is generically distinct: Euclidean continuations removing a
conical singularity can be defined at each horizon separately but not
simultaneously at both \cite{Ginsparg:1982rs} (see also
\cite{Bousso:1996au,Bousso:1997wi}). The only (degenerate) exception
is in an extremal, or Nariai, limit \cite{Nariai} where both
periodicities of Euclidean time match: the spacetime develops a
nearly $dS_2$ throat in this extremal limit \cite{Ginsparg:1982rs}.
More on the nearly $dS_2$ limit and the wavefunction of the universe
appears in \cite{Maldacena:2019cbz} (see also \cite{Anninos:2012ft}).
Related discussions with some
relevance to this paper also appear in \cite{Fernandes:2019ige}.

The general $d+1$-dimensional SdS spacetime is of similar form as
(\ref{SdSst}) but with\ 
$f(r)=1-\frac{2m}{l} (\frac{l}{r})^{d-2}-\frac{r^2}{l^2}$,
and will have qualitative parallels.
We will focus on the 4-dim Schwarzschild de Sitter case in what
follows.\ For $SdS_4$, the function $f(r)$ is a cubic and the zeroes
of $f(r)$, \ie\ solutions to $f(r)=0$, give the locations of the
horizons. We can parametrize this as
\bea\label{SdS4-fmrsrD}
f(r) = 1-\frac{2m}{r}-\frac{r^{2}}{l^{2}} = \frac{1}{l^2\,r}\,
(r_D-r)(r-r_S)(r+r_S+r_D)\,,\qquad\qquad\qquad \nn\\ [1mm]
r_Dr_S(r_D+r_S)=2ml^2\,,\qquad  r_D^2+r_Dr_S+r_S^2=l^2\ ;\qquad
0\leq r_S\leq r_D \leq l\ ;\qquad {m\over l}\leq {1\over 3\sqrt{3}}\ .
\eea
We will take the roots $r_S$ and $r_D$ to label the Schwarzschild
black hole and de Sitter (cosmological) horizons respectively.
(The third zero $-(r_{D}+r_{S})$ does not correspond to a physical horizon.)
The roots $r_S, r_D$ are constrained as above.   

\begin{figure}[h] 
\bc\includegraphics[width=26pc]{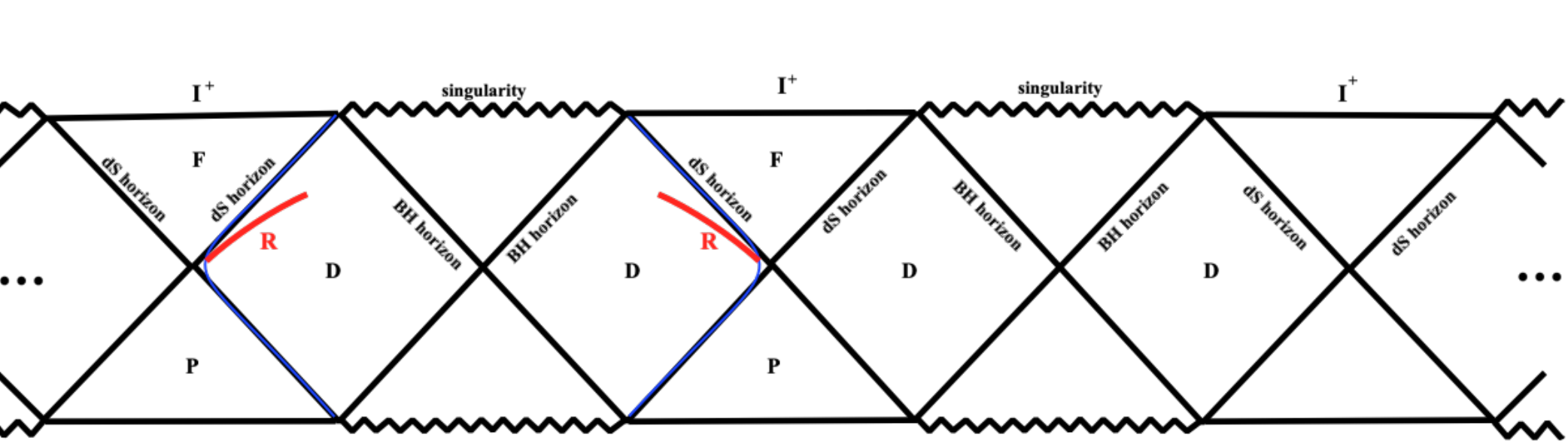}\ec \vspace{-5mm}
\caption{{ \label{figSdSpD}
    \footnotesize{The Schwarzschild de Sitter black hole Penrose diagram, with focus on one black hole ``patch''.
}}}
\end{figure}

The case with $m=0$, or $r_S=0,\ r_D=l$, is pure de Sitter space,
while the flat space Schwarzschild black hole has $l=0$, or
$r_S=m,\ r_D=0$. The above structure of horizons is valid for ${m\over
  l}<{1\over 3\sqrt{3}}$\,, beyond which there are no horizons
\cite{Gibbons:1977mu}.  The limit ${m\over l}={1\over 3\sqrt{3}}$
corresponds to the cosmological and Schwarzschild horizon values
coinciding: here we have $r_S=r_D=r_0={l\over\sqrt{3}}$\ from
(\ref{SdS4-fmrsrD}). This special value leads to the extremal, or
Nariai, limit where the near horizon region (between the horizons)
becomes $dS_2\times S^2$.  Overall the range of physically interesting
$r_S, r_D$ satisfies\ $0 < r_S < r_0 < r_D$ for generic values. The
fact that $r_S<r_D$ implies that the cosmological horizon is
``outside'' the Schwarzschild one.
The black hole interior has $r<r_S$ with $r\ra 0$ the singularity.
The region $r_D<r\leq\infty$ describes the future and past de Sitter
universes, with $r\ra\infty$ the future boundary $I^+$ (or past,
$I^-$).  The maximally extended Penrose diagram Figure~\ref{figSdSpD}
shows an infinitely repeating pattern of Schwarzschild
coordinate patches or ``unit-cells'' containing Schwarzschild black
hole horizons cloaking interior regions: these patches are bounded by
cosmological horizons on the left and right, with future/past
universes beyond the cosmological horizons.

The intermediate static diamond region $D$ is the exterior of the black
hole, \ie\ the static patch with a timelike Killing vector where physical
timelike observers can be stationary:
\be\label{DrSrrD}
D:\qquad r_S < r < r_D\ ;\qquad 0<f(r)<1\ .
\ee
We want to consider the limit of a ``small'' black hole in de Sitter, \ie\
\be\label{m<<l}
m \ll l\,,\qquad\quad l\ra {\rm large}\qquad\Rightarrow\qquad r_D\gg r_S\ .
\ee
The horizon locations can then be found perturbatively to be\
$r_S\simeq 2m\,,\ r_D\simeq l-m \gg r_S$\,,\ from (\ref{SdS4-fmrsrD}).\
In this limit (in a sense opposite to the Nariai limit where
$r_S\sim r_D$), the black hole is much smaller than the ambient
de Sitter scale, \ie\ we have a small black hole in a large accelerating
universe. So we expect that the ambient cosmology can be taken as a
frozen classical background while the black hole undergoes Hawking
evaporation. This is corroborated by the fact that the black hole
Hawking temperature is much larger than the Gibbons-Hawking temperature
of the ambient de Sitter horizon: \ie\ using the surface
gravities $\kappa$\ \cite{Bousso:1996au,Bousso:1997wi} (see also
\cite{Shankaranarayanan:2003ya}) and $T={\kappa\over 2\pi}$
we obtain\footnote{  $\kappa_{BH, dS}={1\over 2\sqrt{f(ml^2)}}
  \big\vert{df\over dr}\big\vert_{r_S,r_D}$
  which give ${1\over 2\beta_{S,D}}\,{1\over\sqrt{1-3(m/l)^{2/3}}}$, with
  $\beta_{S,D}$ in (\ref{beta}).} in the limit (\ref{m<<l}),
\be\label{Tbh>>TdS}
T_{BH}\sim {1\over 8\pi m}\ ,\qquad T_{dS}\sim {1\over 2\pi l}\ ;\qquad\qquad
T_{dS} \ll T_{BH}\ .
\ee
Pushing this to the extreme leads to the flat space limit
\be\label{flatLimit}
r_D\sim l\ra \infty\,:\qquad {r_D\over l} \ra 1\ ,
  \qquad r_S\ra 2m\ ;\qquad\quad T_{dS}\ra 0\ ,
\ee
where the ambient de Sitter background acquires a vanishingly small
temperature, approaching asymptotically flat space. Our entire analysis
will in fact focus on these limits (\ref{m<<l}), (\ref{Tbh>>TdS}),
with the flat limit (\ref{flatLimit}) as a special case to corroborate
with.

Towards analysing the generalized entropy, we will require recasting
the Schwarzschild de Sitter metric (\ref{SdSst}) in Kruskal coordinates
which are regular at the black hole horizon. These are not regular in
the vicinity of the de Sitter horizon (where a distinct set of Kruskal
variables is more appropriate), but we will find that the black hole
Kruskal variables suffices for our considerations. This is consistent
with the fact the ambient de Sitter space simply serves as a frozen
classical background in our regimes of interest. 
With this in mind, we define the black hole tortoise coordinate
following the discussion in \cite{Guven:1990ubi}\,:
\be\label{tortoise coordinate}
r^{\ast}=\int \frac{1}{f(r)} \,dr = \int \frac{1}{1-\frac{2m}{r}-\frac{r^{2}}{l^{2}}} \ dr=\int \frac{l^{2}r}{(r_{D}-r)(r-r_{S})(r+r_{S}+r_{D})} \ dr\ .
\ee
Taking $f(r)>0$ as pertains to the region $D$ in (\ref{DrSrrD}), 
this gives 
\be\label{tortoise solution}
e^{r^{\ast}}=(r_{D}-r)^{-\beta_{D}}(r-r_{S})^{\beta_{s}}(r+r_{D}+r_{S})^{\beta_{M}}\ ,
\ee
with the parameters (which simplify $dr^\ast/dr$ to $1/f(r)$)
\be\label{beta}
\beta_{D}=\frac{l^{2}r_{D}}{(r_{D}-r_{S})(2r_{D}+r_{S})}\,,   \quad
\beta_{S}=\frac{l^{2}r_{S}}{(r_{D}-r_{S})(2r_{S}+r_{D})}\,,\quad
\beta_{M}=\frac{l^{2}(r_{D}+r_{S})}{(2r_{D}+r_{S})(2r_{S}+r_{D})}\,.
\ee
In the flat limit (\ref{flatLimit}), $\beta_S\ra r_S$ and $\beta_D\ra l$.

The $SdS_4$ metric (\ref{SdSst}) is recast 
as $ds^2= f(r)(-dt^2+{dr^\ast}^2)+r^{2}d\Omega_{2}^2$\,.\
In the neighborhood of the black hole horizon, the Kruskal coordinates
are then defined as $U_S,\, V_S$\,, and the Schwarzschild de Sitter metric
becomes \cite{Guven:1990ubi}
\bea\label{SdS-Kruskal}
\alpha_{S}=\frac{1}{2\beta_{S}}\,;\qquad
U_S V_S = - e^{2\al_S r^\ast}\ ,\qquad {U_S\over V_S} = - e^{-2\al_S t}\ ;
\qquad 
ds^{2} = -\frac{dU_{S}dV_{S}}{W^2}+r^{2}d\Omega^{2}\ ,\nn\\ [2mm]
W = \sqrt{r}\,l\,\alpha_{S}(r_{D}-r)^{\frac{-(1+2\alpha_{S}\beta_{D})}{2}}(r-r_{S})^{\frac{2\alpha_{S}\beta_{S}-1}{2}}(r+r_{S}+r_{D})^{\frac{2\alpha_{S}\beta_{M}-1}{2}}\ .
\qquad
\eea
The value of $\al_S$ here ensures regularity at the black hole horizon.
(noting $\beta_M+\beta_S=\beta_D$ we see that $W$ has dimensions of
inverse length.)\ The Kruskal coordinates cover both the left and right
static diamonds $D$ of the black hole patch: in the right side
(containing $R_+$ in Figure~\ref{figSdSpD}) we define
\be\label{SdS-Kruskal2}
U_{S} = -e^{-\alpha_{S}(t-r^{\ast})}\,,\qquad V_{S} = e^{\alpha_{S}(t+r^{\ast})}\ ,
\ee
while on the left side (with $R_-$) there is a relative minus sign,
\ie\ $U_S\ra -U_S,\ V_S\ra -V_S$.

For our purposes, it is a reasonable approximation to look at the
s-wave sector of the black hole and consider the bulk matter as a
2-dim CFT: this enables the use of 2-dim CFT tools to study the
entanglement entropy of bulk matter. With this in mind, we consider
a reduction ansatz of the form 
\be\label{redux+weyl}
ds^2_{(4)} = g^{(2)}_{\mu\nu} dx^\mu dx^\nu + \lambda^{-2} \phi\,d\Omega_2^2\ ;
\qquad\quad g_{\mu\nu}=\phi^{1/2}\,g^{(2)}_{\mu\nu}\ ,
\ee
to obtain 2-dimensional dilaton gravity
\cite{Strominger:1994tn,Grumiller:2002nm} (see also
\cite{Narayan:2020pyj};  applications to certain families of cosmologies
appears in \cite{Bhattacharya:2020qil}). The lengthscale
$\lambda^{-1}$ has been introduced to make the dilaton dimensionless.
The dilaton $\phi$ translates to the 4-dim transverse area of 2-spheres\
$4\pi\phi\over\lambda^2$\,.  
The final term represents a Weyl transformation to absorb the dilaton
kinetic term giving\
${1\over 16\pi G_2}\int d^2x \sqrt{-g}\ (\phi {\cal R}
- {6\over l^2} \phi^{1/2} + 2\lambda^2\phi^{-1/2})$\ as the 2-dim action.\
The 2-dim metric and dilaton are
\be\label{SdS-2d}
ds^{2} = - \lambda\,r \frac{dU_{S}dV_{S}}{W^2}\, \equiv\,
- \frac{dU_{S}dV_{S}}{(W')^2}\ ,
\qquad\qquad \phi=r^2\lambda^2\ ,
\ee
where $W'={W\over \sqrt{\lambda\,r}}$ and $W$ is the conformal factor
given in (\ref{SdS-Kruskal}). For our discussions of entanglement entropy
in these 2-dim theories, $\lambda$ will be regarded as some fixed length
scale independent of the de Sitter scale $l$ so as to not interfere with
the flat limit. With $G_{_N}$ the 4-dim Newton constant,\
$G_2={G_{_N}\over V_2}$ and $V_2={4\pi\over\lambda^2}$\,, the
area term in the 2-dim theory is\
${\phi\over 4G_2}={4\pi\,r^2\over 4G_{_N}}$\ equivalent to the 4-dim one.

\section{Black holes in de Sitter and entanglement entropy}\label{sec:SdSbhEE}

In the regimes (\ref{m<<l}), (\ref{Tbh>>TdS}), that we are
considering, we see that there is a version of the information paradox
for the black hole that is at play, with some parallels with that for
the eternal $AdS$ black hole in \cite{Almheiri:2019yqk}. If the black
hole forms from initial collapsing matter in a pure state, then
information recovery at late times compared with the black hole
timescale requires that the generalized gravitational entropy of the
radiation obeys the Page curve. Note that in the current situation,
this is only approximate since the ambient de Sitter space is only
consistent with bulk CFT matter in a thermal state at the de Sitter
temperature. However if the ambient de Sitter temperature is very low,
then one might imagine that a pure state approximation is
reasonable. This is the limit (\ref{Tbh>>TdS}) we study to understand
the black hole evaporation process here, which we find to be
consistent. We will find, as in various previous investigations, that
a nontrivial island emerges at late times a little outside the horizon
as a nontrivial quantum extremal surface solution to the extremization
of the generalized gravitational entropy. Including this and using the
island rule shows the late time entropy to be bounded. Our
analysis has close parallels with that of flat space Schwarzschild
black holes \eg\ \cite{Anegawa:2020ezn,Hashimoto:2020cas}, our
expressions showing essential agreement in the flat space limit
(\ref{flatLimit}).

In our case, we consider distant observers that are stationary,
represented by timelike worldlines in the static patch $D$ region
(\ref{DrSrrD}), between the Schwarzschild and de Sitter (cosmological)
horizons. Towards simulating the flat space limit, we will consider
the far end of the radiation region as asymptoting to the cosmological
horizon. The outgoing radiation hits the cosmological horizon and as
such is expected to go beyond and eventually hit the future boundary
$I^+$ (future timelike infinity). However we assume that the observers
propagating in the static patch $D$ collect this outgoing radiation.
So we will focus on the Schwarzschild patch bounded by the
cosmological horizons on both sides and ignore the regions beyond: see
Figure~\ref{figSdSqesI}.

The island proposal \cite{Almheiri:2019hni} for the fine-grained entropy
of the Hawking radiation is 
\be\label{Island formula}
S(R)=min\left\{{ext\Big[\frac{Area(\partial I)}{4G_{_N}}
      + S_{matter}(R\cup I)\Big]}\right\}
\ee
where $R$ is a region far from the black hole where the radiation is
collected by distant observers and $I$ is a spatially disconnected
island around the horizon that is entangled with $R$. The intuition
behind this expression is that after about half the black hole has
evaporated, the Hawking radiation going out (roughly $I$) begins to
purify the radiation that was emitted early on (roughly $R$). This
purification of the early radiation by the late Hawking radiation is a
reflection of the entanglement between the two parts, and stems from
the picture of Hawking radiation as due to production of entangled
particle pairs near the horizon (which is taken as vacuum). Thus
$R\cup I$ purifies over time and its entanglement thus does not grow:
the slowly evaporating black hole has decreasing area, so $S(R)$
decreases in time.

The Hawking process is dominated by s-wave modes. So we calculate the
bulk entropy technically using 2-dimensional techniques where we
approximate the bulk matter by a 2-dim CFT propagating in the 2-dim
background obtained by dimensional reduction. In what follows, we will
employ these techniques to calculate the entanglement entropy in the
Schwarzschild de Sitter geometry by considering the 2-dim background
(\ref{SdS-2d}) obtained from the reduction of (\ref{SdS-Kruskal}).
There is no holography in our entire discussion here: we are simply
calculating the entropy of radiation in the bulk spacetime using the
island rule (\ref{Island formula}).

In the absence of the island, the radiation regions are given by the
intervals\ $R_\pm \in D$,
\be\label{radregionR}
R = R_-\cup R_+\ ;\qquad
R_-\equiv [r_{D-}, b_-]\,,\quad R_+\equiv [b_+, r_{D+}]\ ,
\ee
which are the two regions marked in the Figure, one in either
asymptotic region of the black hole patch. Since the radiation region
is far from the black hole horizon, we have\ $b_\pm-r_S \gg r_S$.\ 
The entropy of the Hawking radiation is
\be\label{without island formula}
S_{matter} = S(R)\ .
\ee
In the 2-dim CFT, the matter entanglement entropy for a single
interval $A=[x,y]$ is obtained from the replica formulation
\cite{Calabrese:2004eu,Calabrese:2009qy} 
after also incorporating in $d[x,y]$ the
conformal transformation\footnote{\label{Footnote1} Any 2-dim metric
  is conformally flat so $ds^2=e^f\eta_{\mu\nu}dx^\mu dx^\nu$.
The twist operator 2-point function scales under a conformal
transformation as\
$\lan \sigma(x_1)\,\sigma(x_2)\ran_{e^fg} = e^{-\Delta_n\,f/2}\vert_{x_1}\,
e^{-\Delta_n\,f/2}\vert_{x_2}\, \lan \sigma(x_1)\,\sigma(x_2)\ran_{g}$ with
$\Delta_n={c\over 12} {n^2-1\over n}$.
Since the partition function in the presence of twist operators scales
as the twist operator 2-point function, the entanglement entropy becomes\
$S^{12}_{e^fg} = -\lim_{n\ra 1}\,\del_n \lan \sigma(x_1)\,\sigma(x_2)\ran_{e^fg}
= S^{12}_g + {c\over 6}\sum_{endpoints} \log e^{f/2}$ ,\ 
giving for a single interval\ \
$S^{12}_g = {c\over 6}\log \big({\Delta^2\over\epsilon_{UV}^2}\big)
\ \ra\ 
S^{12}_{e^f g} = {c\over 6}\log \big({\Delta^2\over\epsilon_{UV}^2}\,
e^{f/2}|_1\, e^{f/2}|_2\big)$.
} to a curved space \cite{Almheiri:2019psf},
stemming from the $W'$-factor in the 2-dim metric (\ref{SdS-2d}),
\be\label{single interval}
S_{A}=\frac{c}{3}\log{d[x,y]} \,=\,
{c\over 6} \log \left({-\Delta U_S \Delta V_S\over W'|_x\,W'|_y}\right)\ .
\ee
From (\ref{SdS-2d}), we see that there is one factor of $\lambda$ that
arises inside the logarithm in each expression of the last form: in
addition there is the ultraviolet cutoff $\epsilon_{UV}$\ as in
footnote~\ref{Footnote1}. This factor ${\lambda\over\epsilon_{UV}^2}$
in each such term will not play much role and we will ignore this
except where necessary.

The entanglement entropy for multiple disjoint intervals
\be
A=[x_{1},y_{1}]\ \cup\ [x_{2},y_{2}]\ \cup\ \ ....\ \ \cup\ [x_{N},y_{N}]
\ee
is more complicated: this arises from the multi-point correlation
functions of twist operators and so it depends on not just the
central charge but detailed CFT information. In the limit where the
intervals are well-separated, one can expand the twist operator products
and obtain \cite{Calabrese:2004eu,Calabrese:2009qy,
  Calabrese:2009ez,Calabrese:2010he}
\be\label{multiple interval}
S_A = \frac{c}{3}\,\log\, { \prod_{i,j} d[x_{j},y_{i}] \over \,
  \prod_{i<j} d[x_{j},x_{i}]\ \ \prod_{i<j} d[y_{j},y_{i}]\, }
\ee
For two intervals $[x_1,y_1]\cup [x_2,y_2]$, this is a limit where
the cross ratio $x$ is small, \ie\
\be\label{crossRatio}
x = {\, d[x_1,y_1]\, d[x_2,y_2]\,\over\, d[x_1,x_2]\,d[y_1,y_2]\,}\ ;
\qquad x\ll 1\ ,   
\ee
where we use the Kruskal distances in (\ref{single interval}) in
constructing the cross-ratio.
In 2-dim CFTs with a holographic dual, this is the situation where the
two intervals $A, B$ are well-separated and their mutual information 
exhibits a disentangling transition \cite{Headrick:2010zt} with
$I[A,B]=S[A]+S[B]-S[A\cup B]\ra 0$, \ie\ the disconnected surface
$S_{dis}=S[A]+S[B]$ has lower area than the connected surface
$S_{conn}=S[A\cup B]$. It turns out that the cross-ratio for the points
in question becomes small at late times, as we will see later, justifying
the use of (\ref{multiple interval}) for our purposes.

In what follows, we first calculate the entanglement entropy for the
configuration without any island, using \ref{without island formula}:
this shows the entropy of radiation as increasing linearly in time at
late times, indicative of the information paradox. Towards resolving
this we will include a possible island and calculate the entanglement
entropy using the island rule \ref{Island formula}: this results in a
late time entropy that is time-independent.

\subsection{Entanglement entropy: no island}\label{sec:noIsl}

In this section, we will evaluate the entanglement entropy of the
radiation at late times in the absence of any island. 
Then we have only the radiation regions $R_-\cup R_+$ in
Figure~\ref{figSdSqesI}, given by the intervals (\ref{radregionR}),
\ie\ $R_-\equiv [r_{D-}, b_-]$ and $R_+\equiv [b_+, r_{D+}]$. In the
limit (\ref{m<<l}), (\ref{Tbh>>TdS}), with the de Sitter temperature
very low, we can approximate the entire system as a pure state on any
slice. Then the bulk matter CFT entropy of $R$ is the same as that of
the complementary region $R_c=[b_{-},b_{+}]$, so we obtain
\be\label{S_matter}
    S_{matter} = \frac{c}{3}\,\log\,[d(b_{+},b_{-})]\ .
\ee
We label the spacetime coordinates in the left and right asymptotic
regions in the Schwarzschild patch as
\be\label{b-b+beta}
b_+:\ \ \ (t,r)=(t_{b},b)\ ,\qquad
b_-:\ \ \ (t,r)=(-t_{b}+\frac{i\beta}{2},b)\ ;\qquad\quad
\beta=\frac{2\pi}{\alpha_{S}}\ .
\ee
This choice of $\beta$ is simply a convenient way of incorporating the
relative minus signs in the Kruskal coordinates (\ref{SdS-Kruskal}),
(\ref{SdS-Kruskal2}), in the left and right regions through $e^{i\beta
  \al_S/2} = e^{i\pi} = -1$.  With this parametrization of the left
and right time coordinates, we can conveniently use the expressions in
(\ref{SdS-Kruskal2}), with $\beta$ automatically doing the left-right
book-keeping.

Then we evaluate the bulk matter entropy in the Schwarzschild de Sitter
geometry (\ref{SdS-2d}) using (\ref{single interval}) to obtain
\be\label{matter entropy}
S_{\text{matter}} = \frac{c}{6} \cdot \log \left(b\, \frac{(U_{S}(b_{-})-U_{S}(b_{+}))(V_{S}(b_{+})-V_{S}(b_{-}))}{W(b_{+})W(b_{-})}\right)\ .
\ee
The total entanglement entropy then becomes
\be\label{Total entropy}
S = \frac{c}{6}\,\log \left[16\beta_{S}^{2}\,(b-r_{S})\,
\frac{(r_{D}-b)(b+r_{S}+r_{D})}{l^2} \, \cosh^{2}\Big(\frac{t_{b}}{2\beta_{S}}\Big)\right]\,.
\ee
The details of the calculation are shown in Appendix \ref{App:noIsland}.\
The late time approximation is\
$t_{b}\gg b>r_{S}\,,\ \frac{t_{b}}{r_{S}}\gg 1$\,: the above result then
approximates as
\be\label{approx.entropy}
S\approx const + \frac{c}{6}\,\frac{t_{b}}{\beta_{S}}\ .
\ee
Now in the flat space limit (\ref{flatLimit}), the total entropy at
late times becomes\ $S\approx \frac{c}{6} \frac{t_{b}}{r_{S}}$\,,
in agreement with \cite{Anegawa:2020ezn,Hashimoto:2020cas}.\ 
This linear growth in time means that the entropy of the radiation
will eventually be infinitely larger than the Bekenstein-Hawking
entropy of the black hole.


\section{Late time entanglement entropy with island}\label{sec:EEisland}

In this section we will evaluate the entropy of radiation in the
presence of an island and show that it saves the entropy bound,
recovering the Page curve for the black hole in Schwarzschild de
Sitter space at late times $t_{a},t_{b}\gg r_{S}$.
The island is the region marked $I$ in Figure~\ref{figSdSqesI}: 
the intervals in question are
\be\label{R-UIUR+}
R_-\equiv [r_{D-}, b_-]\,,\qquad R_+\equiv [b_+, r_{D+}]\ ,\qquad
I \equiv [a_-,a_+]\ .
\ee
Since we are considering the limit (\ref{m<<l}), (\ref{Tbh>>TdS}),
with the black hole evaporation well-separated from de Sitter physics,
we expect the island in question to emerge near the black hole horizon.
Then
\be\label{b-rs>>a-rs, b>>rs}
b-r_{S} \gg a-r_{S} \sim 0\ ;\qquad b\gg r_S\ ,
\ee
and the last inequality reflects the fact that we are considering
distant observers far from the black hole.
Since the ambient de Sitter temperature is very low in our
considerations as mentioned earlier, we approximate the bulk matter
to be in a pure state so the entanglement entropy of
$A\equiv R_-\cup I \cup R_+$ is equal to that of the
complementary intervals $A_c \equiv [b_-,a_-]\cup [a_+,b_+]$\ 
(see (\ref{S 3 interval}), for details on the intervals $A$).\ The
assumptions (\ref{b-rs>>a-rs, b>>rs}) imply that the intervals are
well-separated: then we can express the entanglement entropy for $A_c$
using \ref{multiple interval} as
\be\label{S_matter 2}
S_{matter}=\frac{c}{3}\,\log\, \frac{d[a_{+},a_{-}]\, d[b_{+},b_{-}]\,
  d[a_{+},b_{+}]\, d[a_{-},b_{-}]}{d[a_{+},b_{-}]\,d[a_{-},b_{+}]}\ .
\ee
For the intervals $[b_-,a_-]\cup [a_+,b_+]$, as the cross-ratio $x$ in
(\ref{crossRatio}) becomes small, we have
$1-x={\,d[a_{+},a_{-}]\, d[b_{+},b_{-}]\,\over \,d[a_{+},b_{-}]\,d[a_{-},b_{+}]\,}
\ra 1$.
Adding the area term ${\phi\over 4G_2}={4\pi a^2\over 4G_{_N}}$ for both
left and right sides, the total generalized entropy becomes\ $S_{total}\sim 
2 \big({\pi a^2\over G_{_N}} + {c\over 3} \log\,d[a_+,b_+]\big)$ since the
left and right sides give essentially equal contributions. In detail,
using the Kruskal coordinates (\ref{SdS-Kruskal}), (\ref{SdS-Kruskal2}),
(\ref{A1}), we evaluate the total generalized entropy (\ref{S_matter 2})
obtaining
\begin{align}\label{Stotal=area+log.a+a-b+b-.a+b+a-b-/a+b-a-b+}
    S_{total} & =\frac{2\pi a^{2}}{G_{_N}}+\frac{c}{6} \log \Big[\frac{2^8r_{S}^4}{(\frac{r_{D}-r_{S}}{l})^4(\frac{2r_{S}+r_{D}}{l})^4}(a-r_{S})(b-r_{S})(\frac{a-r_{D}}{l})(\frac{b-r_{D}}{l})
    \cdot \nn\\
    & \qquad\qquad\qquad\qquad
    (\frac{a+r_{S}+r_{D}}{l})(\frac{b+r_{S}+r_{D}}{l}) \cosh^2{\frac{t_{a}}{2\beta_{S}}} \cosh^2{\frac{t_{b}}{2\beta_{S}}}\Big]\nn\\
    & \qquad\qquad\qquad + \frac{c}{3} \log \Big[1-2\frac{(a-r_{S})^{\alpha_{S}\beta_{S}}}{(b-r_{S})^{\alpha_{S}\beta_{S}}}\,C(a)\, \cosh{\big(\alpha_{S}(t_{a}-t_{b})\big)}\Big]\nn\\
    & \qquad\qquad\qquad -\frac{c}{3} \log \Big[1+2\frac{(a-r_{S})^{\alpha_{S}\beta_{S}}}{(b-r_{S})^{\alpha_{S}\beta_{S}}}\,C(a)\, \cosh{\big(\alpha_{S}(t_{a}+t_{b})\big)}\Big]\ ,
\end{align}
with $C(a)$ defined as
\be\label{C(a)rDab}
C(a) = \frac{(r_{D}-b)^{\alpha_{S}\beta_{D}}(a+r_{S}+r_{D})^{\alpha_{S}\beta_{M}}}{(r_{D}-a)^{\alpha_{S}\beta_{D}}(b+r_{S}+r_{D})^{\alpha_{S}\beta_{M}}}\ .
\ee
See Appendix~\ref{App:Island} for details of this calculation. At late
times, $t_a, t_b\gg r_S$, we approximate\ $\cosh(\al_S(t_a+t_b))\sim
2\cosh(\al_St_a)\, \cosh(\al_St_b)$ which is large. Then taking into
account the expectation $a-r_S\sim 0$\ and simplifying, we obtain
\begin{align}
       \label{S_total}
       S_{total} & = \frac{2\pi a^2}{G_{_N}}+\frac{c}{6}\log\Big[\frac{16 r_{S}^4 (b-r_{S})^2}{(\frac{r_{D}-r_{S}}{l})^4(\frac{2r_{S}+r_{D}}{l})^4}
\frac{(r_{D}-a)(r_{D}-b)(a+r_{S}+r_{D})(b+r_{S}+r_{D})}{l^4\ (C(a))^2}\Big]        \nn \\ 
       & \qquad\qquad\qquad\qquad\qquad\qquad\
       -\frac{2c}{3} \sqrt{\frac{a-r_{S}}{b-r_{S}}}\, C(a)\, \cosh{\big(\alpha_{S}(t_{a}-t_{b})\big)}\ .
\end{align}
In the flat space limit (\ref{flatLimit}), we take $r_D$ large and expand
using (\ref{beta}) to approximate (\ref{C(a)rDab}) as
\be\label{flatLim-rDab}
C(a) \,\sim\, \Big(1-{b-a\over r_D}\Big)^{\alpha_{S}\beta_{D}}
\Big(1-{b-a\over r_D+r_S}\Big)^{\alpha_{S}\beta_{M}} \,\sim\, 1-{b-a\over 2r_S}\ ,
\ee
which can further approximated as $e^{-{b-a\over 2r_S}}$ in the regime
(\ref{b-rs>>a-rs, b>>rs}) to give the total entropy at late times as
\be\label{flat limit 2}
S_{total}\approx \frac{2\pi a^2}{G_{_N}} + \frac{c}{6} \log\Big[16r_{S}^4(b-r_{S})^2\,e^{{b-a\over r_S}} \Big]-\frac{2c}{3}\sqrt{\frac{a-r_{S}}{b-r_{S}}}\,e^{-{b-a\over 2r_S}}\,\cosh{\big(\alpha_{S}(t_{a}-t_{b})\big)}\ .
\ee
This is in agreement with \eg\ \cite{Hashimoto:2020cas} upto a factor
of $ab$ inside the logarithm, stemming from the fact that we are using
the strict 2-dim bulk metric (\ref{SdS-2d}) after reduction, with the
additional conformal factor in $W'$ relative to $W$ in(\ref{SdS-Kruskal}).
This detailed difference (which also arises in other such expressions)
does not affect the qualitative behaviour of the generalized entropy
in our regime since the s-wave sector is expected to be dominant in
the Hawking process.

Now, extremizing (\ref{S_total}) with respect to the location of the
island boundary $a$ gives
\begin{align}\label{extremize}
&  \frac{\partial S_{total}}{\partial a}=0 \quad   \Rightarrow \\
&  \frac{4 \pi a}{G_{_N}}+ \frac{c}{6}\Big[\frac{(1-2\alpha_{S}\beta_{M})(r_{D}-a)-(1+2\alpha_{S}\beta_{D})(a+r_{S}+r_{D})}{(a+r_{S}+r_{D})(r_{D}-a)}\Big] \nn\\
    & \quad\ 
    -\frac{2c}{3}\cosh{\big(\alpha_{S}(t_{a}-t_{b})\big)} \cdot \sqrt{\frac{a-r_{S}}{b-r_{S}}}\, C(a)\, \Big[\frac{\alpha_{S}\beta_{M}}{a+r_{S}+r_{D}}+\frac{\alpha_{S}\beta_{D}}{r_{D}-a}+\frac{1/2}{a-r_{S}}\Big]=0\ .\nn
\end{align}
Here, since $r_{D}$ is large in our considerations, the second term scales
as $O({1\over r_D})$ and can thus be ignored: further the
${1\over a+r_S+r_D}$ and ${1\over r_D-a}$ also are suppressed relative to
${1\over a-r_S}$\,. With these approximations (\ref{extremize}) becomes
\be\label{simplified extremize}
a \,\simeq\, {1\over \sqrt{a-r_{S}}} \ \frac{G_{_N}c}{12\pi}\,
\frac{1}{\sqrt{b-r_{S}}}\, C(a)\, \cosh{\big(\alpha_{S}(t_{a}-t_{b})\big)} \ .  
\ee 
Now we note that we are in the semiclassical regime where
\be
0 \ll c \ll {1\over G_{_N}}\ ,
\ee
so that the classical area term in the generalized entropy is dominant
but the bulk matter makes nontrivial subleading contributions (which are
not so large as to cause significant backreaction on the classical
geometry).

We are looking for an island with boundary $a\sim r_S$ near the black
hole horizon: this corroborates with the fact that since the entire
right hand side is $O(G_{_N}c)$, in the classical limit $c=0$ we have
$a=r_S$. Thus we can solve the above expression in
perturbation theory setting $a\sim r_S$ to find the first order
correction in $G_{_N}c\ll 1$\,: then schematically we have
\be
a-r_{S} \simeq \frac{K^{2}}{r_{S}^{2}} \frac{1}{b-r_{S}}\ ,\qquad
K=\frac{G_{_N}c}{12\pi} \cosh{(\alpha_{S}(t_{a}-t_{b}))}\,C(r_S)\ ,
\ee
and we finally obtain (with $C(r_S)$ from (\ref{C(a)rDab}) with $a=r_S$)
\be\label{a value}
a\simeq r_{S}+ \frac{(G_{_N}c)^{2}}{144 \pi^{2} r_{S}^2(b-r_{S})}\,
C(r_S)^2\,
\cosh^{2}{\big(\alpha_{S}(t_{a}-t_{b})\big)}\ .
\ee
Using (\ref{flatLim-rDab}) and the comments there, we obtain in the
flat space limit (\ref{flatLimit})
\be
a\simeq r_{S}+ \frac{(G_{_N}c)^{2}}{144 \pi^{2} r_{S}^2(b-r_{S})}\,
e^{r_S-b\over r_S}\,\cosh^{2}{\big(\alpha_{S}(t_{a}-t_{b})\big)}\ ,
\ee
in agreement with \cite{Hashimoto:2020cas}.
With the value of $a$ in (\ref{a value}) the total on-shell
entanglement entropy in (\ref{S_total}) becomes
\begin{align}\label{S_on shell late}
S_{o.s.} & = \frac{2 \pi r_{S}^2}{G_{_N}}+ \frac{c}{6} \log\Big[\frac{16\beta_{s}^4(b-r_{S})^{2}}{l^{4}}\,
{(b+r_{S}+r_{D})^{(1+2\alpha_{S}\beta_{M})}\over (2r_{S}+r_{D})^{2\alpha_{S}\beta_{M}-1}}
\,   { (r_{D}-r_{S})^{1+2\alpha_{S}\beta_{D}}\over (r_{D}-b)^{2\alpha_{S}\beta_{D}-1}}\,
\Big]\nn \\
  & \qquad\qquad\qquad\qquad\
  - \frac{c^{2}G_{_N}}{36\pi r_{S}(b-r_{S})}\,C(r_S)^2\,\cosh^{2}{\big(\alpha_{S}(t_{a}-t_{b})\big)}\ .
\end{align}
Now varying and extremizing the above expression with respect to $t_{a}$,
we obtain $t_{a}=t_{b}$.\ Using this in (\ref{S_on shell late}) gives the
entanglement entropy (keeping only leading terms) to be
\be\label{S(t_a=t_b)}
S_{o.s.} = \frac{2 \pi r_{S}^2}{G_{_N}}+ \frac{c}{6} \log\Big[\frac{16\beta_{s}^4(b-r_{S})^{2}}{l^{4}}\,
{(b+r_{S}+r_{D})^{(1+2\alpha_{S}\beta_{M})}\over (2r_{S}+r_{D})^{2\alpha_{S}\beta_{M}-1}}
\,   { (r_{D}-r_{S})^{1+2\alpha_{S}\beta_{D}}\over (r_{D}-b)^{2\alpha_{S}\beta_{D}-1}}\,
\Big]\ ,
\ee
which is time-independent, stemming from the presence of the island.
The first term is twice the Bekenstein-Hawking entropy of the black hole
and the second term, arising from the bulk entropy of the radiation
region purified by the island, is a constant term not growing in time.
This recovers the expectations on the Page curve for the evaporating
small black hole in de Sitter in the limits we are considering
(\ref{m<<l}), (\ref{Tbh>>TdS}), (\ref{flatLimit}).
In our discussion which is entirely gravitational, it is natural to
take the Planck length as the natural UV scale and set\
$\lambda^{-1}\sim\epsilon_{UV}\sim l_P$\,: then putting back the ${\lambda\over\epsilon_{UV}^2}$ factors gives\
$S_{o.s.}\sim A_s+{c\over 3}\log A_S+{c\over 3}\log{b\over l_P}$\,.

Comparison of the late time entanglement entropy without the island
(\ref{approx.entropy}) and that with the island (\ref{S(t_a=t_b)})
enables an estimate of the Page time when the island configuration
emerges as the preferred quantum extremal surface with lower area.
Making a coarse comparison at the Page time $t_P$, 
\be
\frac{c}{6}\,\frac{t_P}{\beta_{S}} \,\sim\, 2S_{BH}
\qquad\Rightarrow\qquad
t_P \,\sim\, {12\pi\,r_S^2\,\beta_S\over G_{_N}c}\ .
\ee
Beyond this time $t_P$, the no-island configuration (\ref{approx.entropy})
has bigger area and we discard it in favour of the island one
(\ref{S(t_a=t_b)}) (which is a new saddle stemming from replica wormholes)
which does not grow in time. Of course all our analysis is carried out
in a quasi-static approximation with $r_S\sim$ fixed, since $r_S$ is
decreasing very slowly in time as the black hole evaporates.

It is interesting to note that no such island configuration near the
black hole horizon emerges at early times, $t_a, t_b\ll r_S$, when not
much Hawking radiation has gone out: see (\ref{No a>r_S}) in
App.~\ref{App:early time entropy }, where we consider a coarse
approximation with $t_a, t_b\sim 0$. However we might imagine that
there arises a ``vanishing extremal surface'' with the island boundary
$a$ far inside the black hole horizon so $a\ll r_S$. Setting up the
generalized entropy using Kruskal coordinates in the black hole
interior and extremizing in fact reveals such an island in (\ref{a
  value 2}), with a corresponding generalized entropy that is not
significant, vindicating approximate purity in the regimes we are
considering.

We have regarded the ambient de Sitter space as a frozen classical
background, at very low temperature, with regard to the black hole
Hawking process, and found consistency in our island studies and the
2-dim CFT calculations approximating the bulk matter to be in a pure
state.  The calculation in (\ref{S 3 interval}) of the entanglement
entropy of the complementary intervals alongwith the regularization
(\ref{rD-reg}) vindicates the approximate purity of the state in this
regime. If we keep the de Sitter temperature as nonzero, it would
appear that \eg\ (\ref{rD-reg0}) would give rise to a growing
entanglement in time; see \cite{Ageev:2022hqc} for some related
comments.

\section{Discussion}\label{sec:Disc}

We have studied 4-dim ``small'' Schwarzschild de Sitter black holes
(\ref{SdSst}) with mass $m$ and de Sitter scale $l$ in the limit $m\ll
l$ where the de Sitter temperature is very low compared with that of
the black hole (\ref{m<<l}), (\ref{Tbh>>TdS}). In this regime which
has qualitative parallels with a flat space limit (\ref{flatLimit}),
the black hole evaporation process is well-separated from any physics
of de Sitter space which can be regarded as a frozen
background. Strictly speaking the pure state input can only be
approximately true in de Sitter where it is consistent to have bulk
matter in a thermal state at the de Sitter temperature. However in the
limit of very low de Sitter temperature, the ambient de Sitter space
is behaving approximately like a zero temperature bath. In this regime
then, we recover the expectations on the Page curve for the Hawking
radiation in the black hole evaporation process, incorporating
appropriate island contributions, as we have seen.  We are simply
regarding this as a gravitational system of a black hole in de Sitter
space, with no explicit recourse to holography or string theory
(although one might generically expect gravity to be intrinsically
holographic): this is consistent with the island rule via replica
wormholes which does not rely on holography. It is unclear if we can
shed light on questions about microstates here: however perhaps
embedding into some $AdS$ (via a $dS$ bubble with a black hole) may be
a way to approach this in principle.

We focussed on one Kruskal black hole patch in the Penrose diagram
Figure~\ref{figSdSpD} of eternal Schwarzschild de Sitter (which in the
maximal extension comprises a line of alternating black hole and
cosmological patches). This single patch (Figure~\ref{figSdSqesI}) is
in some sense equivalent to the Penrose diagram of the flat space
Schwarzschild black hole embedded in de Sitter, with the cosmological
horizons on both sides serving as asymptotic boundaries (akin to null
infinity in flat space).  Technically we dimensionally reduce the
$SdS_4$ spacetime to 2-dimensions and use 2-dim CFT techniques for
calculating the entanglement entropy of bulk matter approximated as a
2-dim CFT. We saw in sec.~\ref{sec:noIsl} that in the absence of the
island the bulk entropy for the radiation region exhibits unbounded
linear growth, inconsistent at late times when it exceeds the entropy
of the black hole. In sec.~\ref{sec:EEisland} including an appropriate
island as in (\ref{R-UIUR+}), upon extremizing the generalized entropy
incorporating the island rule, we find an island a little outside the
horizon (\ref{a value}) semiclassically\,: the late time entropy
(\ref{S_total}) on-shell becomes (\ref{S_on shell late}) and is twice
the Bekenstein-Hawking entropy of the black hole, plus finite bulk
corrections, in the quasi-static approximation.  In the flat space
limit (\ref{flatLimit}), our expressions are in essential agreement
with those in \cite{Hashimoto:2020cas} (as discussed around
(\ref{flat limit 2})). This suggests that contributions due to
further islands in the other Kruskal patches in our regime are
suppressed (if they exist), perhaps consistent with gravity effectively
being very weak far from the black hole.

We have restricted to observers propagating within the static patch,
with $f(r)<1$, as appropriate for the physics pertaining to the
Hawking evaporation of the black hole.  Let us now imagine considering
observers near the future boundary $I^+$ of the de Sitter patch,
beyond the cosmological horizon. Various studies on quantum extremal
surfaces and cosmologies appear in
\eg\ \cite{Krishnan:2020fer}-\cite{Bousso:2022gth}.  In the present
context, one can study the generalized entropy for such $I^+$
observers as well (see \cite{Fernandes:2019ige} for some discussions
on classical RT/HRT surfaces at $I^+$ in $SdS$, and
\cite{Narayan:2020nsc} and earlier work in $dS$). This reveals quantum
extremal surfaces that are timelike-separated from observers at $I^+$:
there are parallels with studies in the Poincare patch of de Sitter
\cite{Chen:2020tes}, \cite{Goswami:2021ksw}. In $SdS$ however, one
might imagine mapping the radiation region $R_+$ to a corresponding
interval at the future boundary \eg\ by sending out light rays from
$R_+$ to $I^+$ (see Figure~\ref{figSdSqesI}). This suggests that
intervals at $I^+$ should also be able to access information about
black hole evaporation. In particular it might seem possible to find
nontrivial island contributions to analyse the generalized entropy for
observers at $I^+$ to access black hole physics in de Sitter.  It
would be interesting to explore these questions further.

More broadly, in our entire analysis, de Sitter space plays very
little role, although the black hole Kruskal coordinates we employed
do encode detailed aspects of the de Sitter space. Strictly speaking
an intermediate regime where the black hole temperature is comparable
to the de Sitter temperature will require a more detailed analysis of
bulk CFT matter in a mixed state corresponding to the thermal state at
the de Sitter temperature, and would be interesting to study as a
nontrivial nonequilibrium situation. The regions in Schwarzschild de
Sitter parameter space we have explored are very far from the Nariai
(extremal) limit where a $dS_2$ throat emerges: see
\eg\ \cite{Kames-King:2021etp,Moitra:2022glw,Svesko:2022txo,Levine:2022wos}
for some recent investigations on the latter. Our island solution
(\ref{a value}) emerges as a self-consistent solution near the black
hole horizon in the regime of very low de Sitter temperature: it would
seem that the extremization equation (\ref{extremize}) has other
solutions as well, pertaining to other regimes of more cosmological
relevance, worth exploring. These might dovetail with various studies
on quantum extremal surfaces and cosmologies
\eg\ \cite{Krishnan:2020fer}-\cite{Bousso:2022gth}, and perhaps
broader issues with de Sitter space
\eg\ \cite{Chandrasekaran:2022cip}.

\vspace{5mm}

{\footnotesize \noindent {\bf Acknowledgements:}\ \ It is a pleasure
  to thank Nori Iizuka, Dileep Jatkar, Alok Laddha, Sandip Trivedi
  and Amitabh Virmani for discussions and comments on a draft.
  This work is partially supported by a grant to CMI from the Infosys
  Foundation.}

\vspace{2mm}


\appendix

\renewcommand{\theequation}{\thesection.\arabic{equation}}

\section{Details: entropy in the no-island case}\label{App:noIsland}

This section contains some details on the calculations of entanglement
entropy in the absence of the island in sec.~\ref{sec:noIsl}.\
Using (\ref{tortoise solution}), (\ref{SdS-Kruskal}), (\ref{SdS-Kruskal2}),
the black hole patch Kruskal coordinates are:
\begin{align}\label{A1}
& U_{S}= -e^{-\alpha_{S}t} (r_{D}-r)^{-\alpha_{S}\beta_{D}} (r-r_{S})^{\alpha_{S}\beta_{S}} (r+r_{D}+r_{S})^{\alpha_{S}\beta_{M}} ,\\
& \label{A2}V_{S}=e^{\alpha_{S}t} (r_{D}-r)^{-\alpha_{S}\beta_{D}} (r-r_{S})^{\alpha_{S}\beta_{S}} (r+r_{D}+r_{S})^{\alpha_{S}\beta_{M}}\ .
\end{align}
Calculating each part of $S_{matter}$ in equation \ref{matter entropy}
separately gives
\be\label{A3}
U_{S}(b_{-})-U_{S}(b_{+})= (r_{D}-b)^{-\alpha_{S}\beta_{D}} (b-r_{S})^{\alpha_{S}\beta_{S}} (b+r_{D}+r_{S})^{\alpha_{S}\beta_{M}} [e^{-\alpha_{S}t_{b}}-e^{\alpha_{S}(t_{b}-\frac{i\beta}{2})}]\ ,
\ee
\be\label{A4}
V_{S}(b_{+})-V_{S}(b_{-})= (r_{D}-b)^{-\alpha_{S}\beta_{D}} (b-r_{S})^{\alpha_{S}\beta_{S}} (b+r_{D}+r_{S})^{\alpha_{S}\beta_{M}} [e^{\alpha_{S}t_{b}}-e^{-\alpha_{S}(t_{b}-\frac{i\beta}{2})}]\ ,
\ee
\be\label{A5}
W(b_{+})=W(b_{-})= \sqrt{b}l\alpha_{S}(r_{D}-b)^{-\frac{1+2\alpha_{S}\beta_{D}}{2}}(b-r_{S})^{\frac{2\alpha_{S}\beta_{S}-1}{2}} (b+r_{S}+r_{D})^{\frac{2\alpha_{S}\beta_{M}-1}{2}}\ ,
\ee
\be\label{A6}
W(b_{+})W(b_{-})=bl^{2}\alpha_{S}^2 (r_{D}-b)^{-(1+2\alpha_{S}\beta_{D})} (b+r_{S}+r_{D})^{(2\alpha_{S}\beta_{M}-1)} (b-r_{S})^{(2\alpha_{S}\beta_{S}-1)}\ .
\ee
Plugging all these expressions together in (\ref{matter entropy}) gives
\begin{align}
    S_{matter} & =  \frac{c}{6} \log\Big[b(r_{D}-b)^{-2\alpha_{S}\beta_{D}}(b-r_{S})^{2\alpha_{S}\beta_{S}}(b+r_{D}+r_{S})^{2\alpha_{S}\beta_{M}} (e^{-\alpha_{S}t_{b}}-e^{\alpha_{S}(t_{b}-\frac{i\alpha_{S}\beta}{2})}) \cdot \nn \\ 
     & ~~  (e^{\alpha_{S}t_{b}}-e^{-\alpha_{S}(t_{b}-\frac{i\alpha_{S}\beta}{2})}) \frac{1}{bl^{2}\alpha_{S}^{2}} (r_{D}-b)^{(1+2\alpha_{S}\beta_{D})} (b-r_{S})^{(1-2\alpha_{S}\beta_{S})} (b+r_{S}+r_{D})^{(1-2\alpha_{S}\beta_{M})}\Big] \nn \\
     & = \frac{c}{6} \log\Big[(r_{D}-b)(b-r_{S})(b+r_{S}+r_{D})\frac{1}{l^{2}\alpha_{S}^2} \Big(2-(e^{(2\alpha_{S}t_{b}-\frac{i\alpha_{S}\beta}{2})}+e^{-(2\alpha_{S}t_{b}-\frac{i\alpha_{S}\beta}{2})})\Big)\Big] \nn \\
     & =  \frac{c}{6} \log\Big[(b-r_{D})(b-r_{S})(b+r_{S}+r_{D})\frac{1}{l^{2}\alpha_{S}^2}\Big(e^{(\frac{2t_{b}}{2\beta_{S}}-i\pi)}+e^{-(\frac{2t_{b}}{2\beta_{S}}-i\pi)}-2\Big)\Big]\ , \label{A7}
\end{align}
using $\beta=\frac{2\pi}{\alpha_{S}}$ from (\ref{b-b+beta}). From
(\ref{SdS-Kruskal}) we have $\al_S={1\over 2\beta_S}$ so this becomes
\be 
S = \frac{c}{6} \log\Big[(b-r_{S}) 4\beta_{S}^{2}\, (\frac{r_{D}}{l}-\frac{b}{l})(\frac{b}{l}+\frac{r_{S}}{l}+\frac{r_{D}}{l})\, 4\cosh^{2}\frac{t_{b}}{2\beta_{S}}\Big]
\ee 
Thus finally, we obtain (\ref{Total entropy}).

\section{Details: late-time entropy with island}\label{App:Island}

Here we give details on sec.~\ref{sec:EEisland}.
We are looking to calculate (\ref{S_matter 2}), \ie\
\be
    S_{matter}=\frac{c}{3}\log \frac{d(a_{+},a_{-})d(b_{+},b_{-})d(a_{+},b_{+})d(a_{-},b_{-})}{d(a_{+},b_{-})d(a_{-},b_{+})}\ .
\ee
Now calculating each part in $S_{matter}$ separately,\\
\be\label{B.37}
    \log[d(a_{+},a_{-})]=\frac{1}{2}\log \frac{[(U_{S}(a_{-})-U_{S}(a_{+}))(V_{S}(a_{+})-V_{S}(a_{-}))]}{W^{\prime}(a_{+})W^{\prime}(a_{-})}
\ee
with $W'$ as in (\ref{SdS-2d}).    
Then we have
\be
U_{S}(a_{-})-U_{S}(a_{+})=(r_{D}-a)^{-\alpha_{S}\beta_{D}}(a-r_{S})^{\alpha_{S}-\beta_{s}}(a+r_{S}+r_{D})^{\alpha_{S}\beta_{M}}[e^{-\alpha_{S}t_{a}}-e^{\alpha_{S}t_{a}} \cdot e^{-\frac{i\alpha_{S}\beta}{2}}]\ ,
\ee
\be
V_{S}(a_{+})-V_{S}(a_{-})=(r_{D}-a)^{-\alpha_{S}\beta_{D}}(a-r_{S})^{\alpha_{S}-\beta_{S}}(a+r_{S}+r_{D})^{\alpha_{S}\beta_{M}}[e^{\alpha_{S}t_{a}}-e^{-\alpha_{S}t_{a}} \cdot e^{\frac{i\alpha_{S}\beta}{2}}]\ ,
\ee
\be
W^{\prime}(a_{+})W^{\prime}(a_{-})=l^{2}\alpha_{S}^2 (r_{D}-a)^{-(1+2\alpha_{S}\beta_{D})} (a+r_{S}+r_{D})^{(2\alpha_{S}\beta_{M}-1)} (a-r_{S})^{(2\alpha_{S}\beta_{S}-1)}\ .
\ee
Putting all these expressions together in (\ref{B.37}) gives
\be\label{B.38}
\log[d(a_{+},a_{-})]=\frac{1}{2}\log \Big[\frac{1}{l^2\alpha_{S}^2}(a-r_{D})(a-r_{S})(a+r_{S}+r_{D})\Big(e^{(2\alpha_{S}t_{a}-\frac{i\alpha_{S}\beta}{2})}+e^{-(2\alpha_{S}t_{a}-\frac{i\alpha_{S}\beta}{2})}-2\Big)\Big]
\ee
Similarly we obtain
\begin{align}\label{B.39}
    \log[d(b_{+},b_{-})] & =\frac{1}{2}\log \frac{[(U_{S}(b_{-})-U_{S}(b_{+}))(V_{S}(b_{+})-V_{S}(b_{-}))]}{W^{\prime}(b_{+})W^{\prime}(b_{-})} \nn \\
    & =\frac{1}{2}\log \Big[\frac{1}{l^2\alpha_{S}^2}(b-r_{D})(b-r_{S})(b+r_{S}+r_{D})\Big(e^{(2\alpha_{S}t_{b}-\frac{i\alpha_{S}\beta}{2})}+e^{-(2\alpha_{S}t_{b}-\frac{i\alpha_{S}\beta}{2})}-2\Big)\Big]
\end{align}
Now, putting (\ref{B.38}) and (\ref{B.39}) together gives,
using $\beta=\frac{2\pi}{\alpha_{S}}$\,,
\begin{align}\label{log.a+a-b+b-}
   {c\over 3} \log[d(a_{+},a_{-})d(b_{+},b_{-})] 
    & = \frac{c}{6} \log\Big[\frac{2^8r_{S}^4}{(\frac{r_{D}-r_{S}}{l})^4(\frac{2r_{S}+r_{D}}{l})^4}(a-r_{S})(b-r_{S})(\frac{r_{D}-a}{l})(\frac{r_{D}-b}{l})
    \cdot \nn\\
    &  \qquad\qquad
    (\frac{a+r_{S}+r_{D}}{l})(\frac{b+r_{S}+r_{D}}{l}) \cosh^2{\frac{t_{a}}{2\beta_{S}}} \cosh^2{\frac{t_{b}}{2\beta_{S}}}\Big]\ .
\end{align}
We next calculate other relevant contributions:
\begin{align}\label{B.41}
    d(a_{+},b_{+})& =\frac{1}{W^{\prime}(a_{+})W^{\prime}(b_{+})}\Big[(U_{S}(b_{+})-U_{S}(a_{+}))(V_{S}(a_{+})-V_{S}(b_{+}))\Big]^{\frac{1}{2}} \nn \\
    & = \frac{1}{W^{\prime}(a_{+})W^{\prime}(b_{+})}\Bigg[2e^{\alpha_{S}(r^{\ast}(a)+r^{\ast}(b))}
    \cdot \nn\\
    & ~~ \Bigg(\cosh{\Big(\alpha_{S}(r^{\ast}(a)-r^{\ast}(b))}\Big)-\cosh{\Big(\alpha_{S}(t_{a}-t_{b})\Big)}\Bigg)\Bigg]^{\frac{1}{2}}
\end{align}

\begin{align}\label{B.42}
    d(a_{-},b_{-})& =\frac{1}{W^{\prime}(a_{-})W^{\prime}(b_{-})}\Big[(U_{S}(b_{-})-U_{S}(a_{-}))(V_{S}(a_{-})-V_{S}(b_{-}))\Big]^{\frac{1}{2}} \nn \\
    & = \frac{1}{W^{\prime}(a_{-})W^{\prime}(b_{-})}\Bigg[2e^{\alpha_{S}(r^{\ast}(a)+r^{\ast}(b))}
    \cdot \nn\\
    & ~~ \Bigg(\cosh{\Big(\alpha_{S}(r^{\ast}(a)-r^{\ast}(b))}\Big)-\cosh{\Big(\alpha_{S}(t_{a}-t_{b})\Big)}\Bigg)\Bigg]^{\frac{1}{2}}
\end{align}
Now, putting (\ref{B.41}) and (\ref{B.42}) together
\begin{align}\label{B.43}
    d(a_{+},b_{+})d(a_{-},b_{-})&=\frac{1}{W^{\prime}(a_{+})W^{\prime}(b_{+})W^{\prime}(a_{-})W^{\prime}(b_{-})}\Bigg[2e^{\alpha_{S}(r^{\ast}(a)+r^{\ast}(b))}
    \cdot \nn\\
    & ~~ \Bigg(\cosh{\Big(\alpha_{S}(r^{\ast}(a)-r^{\ast}(b))}\Big)-\cosh{\Big(\alpha_{S}(t_{a}-t_{b})\Big)}\Bigg)\Bigg]
\end{align}
Similarly
\begin{align}\label{B.44}
    d(a_{+},b_{-})& =\frac{1}{W^{\prime}(a_{+})W^{\prime}(b_{-})}\Big[(U_{S}(b_{-})-U_{S}(a_{+}))(V_{S}(a_{+})-V_{S}(b_{-}))\Big]^{\frac{1}{2}} \nn \\
    & = \frac{1}{W^{\prime}(a_{+})W^{\prime}(b_{-})}\Bigg[2e^{\alpha_{S}(r^{\ast}(a)+r^{\ast}(b))}
    \cdot \nn\\
    & ~~ \Bigg(\cosh{\Big(\alpha_{S}(r^{\ast}(a)-r^{\ast}(b))}\Big)-\cosh{\Big(\alpha_{S}(t_{a}+t_{b}-\frac{i\beta}{2})\Big)}\Bigg)\Bigg]^{\frac{1}{2}}
\end{align}
\begin{align}\label{B.45}
    d(a_{-},b_{+})& =\frac{1}{W^{\prime}(a_{-})W^{\prime}(b_{+})}\Big[(U_{S}(b_{+})-U_{S}(a_{-}))(V_{S}(a_{-})-V_{S}(b_{+}))\Big]^{\frac{1}{2}} \nn \\
    & = \frac{1}{W^{\prime}(b_{-})W^{\prime}(a_{+})}\Bigg[2e^{\alpha_{S}(r^{\ast}(a)+r^{\ast}(b))}
    \cdot \nn\\
    & ~~ \Bigg(\cosh{\Big(\alpha_{S}(r^{\ast}(a)-r^{\ast}(b))}\Big)-\cosh{\Big(\alpha_{S}(t_{a}+t_{b}-\frac{i\beta}{2})\Big)}\Bigg)\Bigg]^{\frac{1}{2}}
\end{align}
Now, putting (\ref{B.44}) and (\ref{B.45}) together
\begin{align}\label{B.46}
    d(a_{+},b_{-})d(a_{-},b_{+}) & = \frac{1}{W^{\prime}(a_{+})W^{\prime}(b_{-})W^{\prime}(a_{-})W^{\prime}(b_{+})} \Bigg[2e^{\alpha_{S}(r^{\ast}(a)+r^{\ast}(b))}
    \cdot \nn\\
    & ~~ \Bigg(\cosh{\Big(\alpha_{S}(r^{\ast}(a)-r^{\ast}(b))\Big)}-\cosh{\Big(\alpha_{S}(t_{a}+t_{b}-\frac{i\beta}{2})\Big)}\Bigg)\Bigg]
\end{align}
Putting (\ref{B.43}) and (\ref{B.46}) together we get
\begin{align}\label{B.47}
    \frac{c}{3} \log \frac{d(a_{+},b_{+})d(a_{-},b_{-})}{d(a_{+},b_{-})d(a_{-},b_{+})} & =\frac{c}{3} \log \Big[\frac{\cosh{\Big(\alpha_{S}(r^{\ast}(a)-r^{\ast}(b))\Big)}-\cosh{\Big(\alpha_{S}(t_{a}-t_{b})\Big)}}{\cosh{\Big(\alpha_{S}(r^{\ast}(a)-r^{\ast}(b))\Big)}-\cosh{\Big(\alpha_{S}(t_{a}+t_{b}-\frac{i\beta}{2})\Big)}}\Big]
\end{align}
Here 
\be\label{B.48}
\cosh{\Big(\alpha_{S}(t_{a}+t_{b}-\frac{i\beta}{2})\Big)}=-\cosh{\Big(\alpha_{S}(t_{a}+t_{b})\Big)}
\ee
and
\begin{align}\label{B.49}
    \cosh{\Big(\alpha_{S}(r^{\ast}(a)-r^{\ast}(b))\Big)} & =\,\frac{1}{2}\,\Big[\frac{(a-r_{S})^{\alpha_{S}\beta_{S}}(a+r_{D}+r_{S})^{\alpha_{S}\beta_{M}}(r_{D}-b)^{\alpha_{S}\beta_{D}}}{(r_{D}-a)^{\alpha_{S}\beta_{D}}(b+r_{D}+r_{S})^{\alpha_{S}\beta_{M}}(b-r_{S})^{\alpha_{S}\beta_{S}}}
    \nn\\
    & \qquad + \frac{(r_{D}-a)^{\alpha_{S}\beta_{D}}(b+r_{D}+r_{S})^{\alpha_{S}\beta_{M}}(b-r_{S})^{\alpha_{S}\beta_{S}}}{(a-r_{S})^{\alpha_{S}\beta_{S}}(a+r_{D}+r_{S})^{\alpha_{S}\beta_{M}}(r_{D}-b)^{\alpha_{S}\beta_{D}}}\Big] \nn\\
    &\, \sim \frac{1}{2}\,
    \frac{(b-r_{S})^{\alpha_{S}\beta_{S}}}{(a-r_{S})^{\alpha_{S}\beta_{S}}}\,{1\over C(a)}\ ,
\end{align}
using the approximations (\ref{b-rs>>a-rs, b>>rs}), and (\ref{C(a)rDab}).
Thus we obtain
\begin{align}\label{log.a+b+a-b-/a+b-a-b+}
\frac{c}{3} \log\frac{d(a_{+},b_{+})d(a_{-},b_{-})}{d(a_{+},b_{-})d(a_{-},b_{+})}\,
  &\, =\,\frac{c}{3} \log\Big[1-2\frac{(a-r_{S})^{\alpha_{S}\beta_{S}}}{(b-r_{S})^{\alpha_{S}\beta_{S}}}\, C(a)\, \cosh{\Big(\alpha_{S}(t_{a}-t_{b})\Big)}\Big]\nn\\
&\quad\ -\frac{c}{3} \log\Big[1+2\frac{(a-r_{S})^{\alpha_{S}\beta_{S}}}{(b-r_{S})^{\alpha_{S}\beta_{S}}}\, C(a)\, \cosh{\Big(\alpha_{S}(t_{a}+t_{b})\Big)}\Big]\ .
\end{align}
The total bulk matter entanglement entropy thus is (\ref{log.a+a-b+b-})
plus (\ref{log.a+b+a-b-/a+b-a-b+}): along with the area term this gives
(\ref{Stotal=area+log.a+a-b+b-.a+b+a-b-/a+b-a-b+}).\ 
At late times, \ie\ $t_{a},t_{b}\gg r_{S}$ the total entanglement entropy
$S_{total}$, after adding the area term, becomes
\begin{align}
    S_{total} & \sim\ \frac{2\pi a^2}{G_{_N}} + \frac{2c}{6} \log\Big[\frac{2^4 r_{S}^2}{(\frac{r_{D}-r_{S}}{l})^2(\frac{2r_{S}+r_{D}}{l})^2}\sqrt{(a-r_{S})(b-r_{S})}\cdot \nn\\
    & \qquad\qquad
    \sqrt{\frac{(r_{D}-a)}{l}\frac{(r_{D}-b)}{l}\frac{(a+r_{S}+r_{D})}{l}\frac{(b+r_{S}+r_{D})}{l}} \cosh{\frac{t_{a}}{2\beta_{S}}} \cosh{\frac{t_{b}}{2\beta_{S}}}\Big]\nn\\
    & \qquad\qquad\qquad\qquad
    +\frac{c}{3} \log\Big[1- 2\frac{(a-r_{S})^{\alpha_{S}\beta_{S}}}{(b-r_{S})^{\alpha_{S}\beta_{S}}}\, C(a)\, \cosh{\Big(\alpha_{S}(t_{a}-t_{b})\Big)}\Big]\nn\\
    & \qquad\qquad\qquad\qquad
    -\frac{c}{3} \log\Big[4\frac{(a-r_{S})^{\alpha_{S}\beta_{S}}}{(b-r_{S})^{\alpha_{S}\beta_{S}}}\, C(a)\,
      \cosh{\frac{t_{a}}{2\beta_{S}}} \cosh{\frac{t_{b}}{2\beta_{S}}}\Big]\ ,
\end{align}
which upon simplifying (taking $a-r_S\sim 0$ so $\log(1-y)\sim -y$)
gives (\ref{S_total}).

\bigskip\bigskip\bigskip  


\noindent \underline{{\bf  Entanglement entropy of the intervals
    $R_-\cup I\cup R_+$}\,:}\label{App:R-IR+}\ \ 
It is instructive to compare the above late time calculation of
the entanglement entropy of the intervals $[b_-,a_-]\cup [a_+,b_+]$ with
that of the original three intervals $R_-\cup I\cup R_+\equiv
[r_{D-},b_{-}]\cup [a_{-},a_{+}]\cup [b_{+},r_{D+}]$ which are
complementary intervals in the black hole Kruskal patch of $SdS$.
Here, $r_{D-}$ and $r_{D+}$ are the boundaries of the entanglement
wedge of the Hawking radiation in the left and right universes respectively.
We will take these two points $r_{D-}$ and $r_{D+}$ to be very close to
the corresponding cosmological (de Sitter) horizons which might be
approximated as the effective boundaries of the black hole Kruskal
patch in the flat space like limit (\ref{flatLimit}) of de Sitter
that we are considering here: see Figure~\ref{figSdSqesI}. Thus we 
define the spacetime coordinates of these two points as\
\be
r_{D+}\,:\quad (t,r)=(t_{D},r_{D}-\delta)\ ;\qquad
r_{D-}\,:\quad (t,r)=(-t_{D}+\frac{i \beta}{2},r_{D}-\delta)\ .
\ee
Under the assumptions (\ref{b-rs>>a-rs, b>>rs}), using
(\ref{multiple interval}), we approximate the entanglement entropy
for $R_-\cup I\cup R_+$ as
\begin{align}\label{S 3 interval}
  S_{matter} & =\frac{c}{3} \log\Big[d(a_{+},a_{-}) d(b_{+},b_{-})\Big] + \frac{c}{3} \log\Big[ \frac{d(a_{+},b_{+})d(a_{-},b_{-})}{d(a_{+},b_{-})d(a_{-},b_{+})}\Big] \nn\\
 & + \frac{c}{3} \log\Big[ \frac{d(a_{+},r_{D-})d(a_{-},r_{D+})}{d(a_{-},r_{D-})d(a_{+},r_{D+})}\Big]
  + \frac{c}{3} \log\Big[ \frac{d(b_{+},r_{D+})d(b_{-},r_{D-})}{d(b_{+},r_{D-})d(b_{-},r_{D+})}\Big] \nn\\ & + \frac{c}{3} \log\Big[d(r_{D+},r_{D-})\Big]\ .
\end{align}
The expressions in the first line are the same as the matter
entanglement entropy for the intervals $[b_-,a_-]\cup [a_+,b_+]$
complementary to $R_-\cup I\cup R_+$. 
Simplifying using the various Kruskal variable distances as described
earlier, we obtain for the expressions in the second and third lines
\begin{align}\label{S total original}
    &  \frac{c}{3} \log\Big[1+2 \frac{\delta^{\alpha_{S}\beta_{D}}(a-r_{S})^{\alpha_{S}\beta_{S}}(a+r_{S}+r_{D})^{\alpha_{S}\beta_{M}}}{(r_{D}-\delta -r_{S})^{\alpha_{S}\beta_{S}}(r_{D}-a)^{\alpha_{S}\beta_{D}}(2r_{D}-\delta +r_{S})^{\alpha_{S}\beta_{M}}}\cosh{\Big(\alpha_{S}(t_{a}+t_{D})\Big)}\Big]\nn\\
    & - \frac{c}{3} \log\Big[1-2 \frac{\delta^{\alpha_{S}\beta_{D}}(a-r_{S})^{\alpha_{S}\beta_{S}}(a+r_{S}+r_{D})^{\alpha_{S}\beta_{M}}}{(r_{D}-\delta -r_{S})^{\alpha_{S}\beta_{S}}(r_{D}-a)^{\alpha_{S}\beta_{D}}(2r_{D}-\delta +r_{S})^{\alpha_{S}\beta_{M}}}\cosh{\Big(\alpha_{S}(t_{a}-t_{D})\Big)}\Big]\nn\\
    & + \frac{c}{3} \log\Big[1-2 \frac{\delta^{\alpha_{S}\beta_{D}}(b-r_{S})^{\alpha_{S}\beta_{S}}(b+r_{S}+r_{D})^{\alpha_{S}\beta_{M}}}{(r_{D}-\delta -r_{S})^{\alpha_{S}\beta_{S}}(r_{D}-b)^{\alpha_{S}\beta_{D}}(2r_{D}-\delta +r_{S})^{\alpha_{S}\beta_{M}}}\cosh{\Big(\alpha_{S}(t_{a}-t_{D})\Big)}\Big]\nn\\
    & - \frac{c}{3} \log\Big[1+2 \frac{\delta^{\alpha_{S}\beta_{D}}(b-r_{S})^{\alpha_{S}\beta_{S}}(b+r_{S}+r_{D})^{\alpha_{S}\beta_{M}}}{(r_{D}-\delta -r_{S})^{\alpha_{S}\beta_{S}}(r_{D}-b)^{\alpha_{S}\beta_{D}}(2r_{D}-\delta +r_{S})^{\alpha_{S}\beta_{M}}}\cosh{(\alpha_{S}(t_{a}+t_{D}))}\Big]\nn\\
    & + \frac{c}{6} \log\Big[\frac{\delta}{\alpha_{S}^2}\frac{(r_{D}-r_{S}-\delta)}{l}\frac{(2r_{D}+r_{S}-\delta)}{l}\cosh^{2}{\frac{t_{D}}{2\beta_{S}}}\Big]\ .
\end{align}
Since $\delta\ll r_D$ and the exponent $\al_S\beta_D\sim {r_D\over r_S}\gg 1$\,,
we see that the the expressions in the first four lines are of the form
$\log (1+\cdots)$ and thus vanishingly small. The last term 
\be\label{rD-reg0}
\frac{c}{3} \log\Big[d(r_{D+},r_{D-})\Big] \sim \frac{c}{6}
\log\Big[\frac{(r_{D}-r_{S})}{l}\frac{(2r_{D}+r_{S})}{l}\,
  \Big(\frac{\delta}{\alpha_{S}^2}\cosh^{2}{\frac{t_{D}}{2\beta_{S}}}\Big)\Big]
\ee
requires a regularization of the cosmological horizon which is akin to
spatial infinity $\iota^0$ in the flat space limit. Let us define
\be\label{rD-reg}
\frac{(r_{D}-r_{S})}{l}\frac{(2r_{D}+r_{S})}{l}\,
     {\delta\,\lambda\over\al_S^2\,\epsilon_{UV}^2}
     \cosh^{2}{\frac{t_{D}}{2\beta_{S}}}\ \ra\ c_D = finite
\ee
where we have reinstated the various lengthscales, as discussed
after (\ref{single interval}), and using (\ref{SdS-2d}).

With this regularization of the observer near the cosmological horizon
$r_D$, we see that the entanglement entropy of the intervals
$[b_-,a_-]\cup [a_+,b_+]$ used in the text and that of the 
complementary intervals $R_-\cup I\cup R_+$ are essentially equivalent,
as expected for bulk matter in a pure state on the entire slice\
$R_-\cup [b_-,a_-]\cup I\cup [a_+,b_+] \cup R_+$\ in the black hole
Kruskal patch in Figure~\ref{figSdSqesI}, in the regime of very low
de Sitter temperature (\ref{m<<l}), (\ref{Tbh>>TdS}), (\ref{flatLimit}).

\section{Entanglement entropy with island at early times}\label{App:early time entropy }

In this section, we study the entanglement entropy looking for an 
island configuration at early times \ie\ at some small $t_{a},t_{b}$\ (with
$t_{a},t_{b}\ll r_{S}$). In this case Hawking quanta have not had time
to escape out so we do not expect any island near the black hole horizon
of the form (\ref{a value}). Indeed, simplifying
(\ref{Stotal=area+log.a+a-b+b-.a+b+a-b-/a+b-a-b+}) at early times
with the coarse approximation $t_a, t_b\sim 0$ gives
\begin{align}\label{S early >}
    S_{total} & =\frac{2\pi a^{2}}{G_{_N}}
    +\frac{c}{6}\log\Big[\frac{2^8r_{S}^4}{(\frac{r_{D}-r_{S}}{l})^4(\frac{2r_{S}+r_{D}}{l})^4}(a-r_{S})(b-r_{S})(\frac{r_{D}-a}{l})(\frac{r_{D}-b}{l})\cdot \nn\\
    & ~~\qquad\qquad\qquad\qquad\qquad\qquad\qquad\qquad
    (\frac{a+r_{S}+r_{D}}{l})(\frac{b+r_{S}+r_{D}}{l})\Big]\nn\\
    &\qquad +\,\frac{c}{3} \log\Big[1-2\frac{(a-r_{S})^{\alpha_{S}\beta_{S}}}
        {(b-r_{S})^{\alpha_{S}\beta_{S}}}\,C(a)\Big]\, 
     -\,\frac{c}{3} \log\Big[1+2\frac{(a-r_{S})^{\alpha_{S}\beta_{S}}}
      {(b-r_{S})^{\alpha_{S}\beta_{S}}}\,C(a) \Big]\ .
\end{align}
Extremizing with $a-r_S\sim 0$ and simplifying with $r_D$ large (so
$C(a)\sim 1$) gives
\be\label{No a>r_S}
         {4\pi a\over G_{_N}} + {c\over 6}\,{1\over a-r_S}
         - {c\over 3}\,{1\over\sqrt{(a-r_S)(b-r_S)}} \sim 0\ ,
\ee
keeping only leading terms. So, since $b\gg r_S$, there is no island
solution with $a\gtrsim r_S$.

However we might imagine that there arises a vanishing extremal surface
with the island boundary $a$ far inside the black hole horizon so
$a\ll r_S$. The boundary $b$ of the entanglement wedge of the Hawking
radiation continues to be far away from the horizon \ie\
$b-r_{S} \gg r_{S}-a$.
Towards analysing this, we will employ Kruskal coordinates different
from those previously used, defined in the black hole interior since
$a< r_S$. So we define the tortoise coordinate as
\be\label{tortoise-In}
r^{\ast}=-\int \frac{1}{f(r)} \,dr = -\int \frac{1}{1-\frac{2m}{r}-\frac{r^{2}}{l^{2}}} \,dr=\int \frac{l^{2}r}{(r_{D}-r)(r_{S}-r)(r+r_{S}+r_{D})} \,dr
\ee
This gives
\be\label{tortoise-In2}
e^{r^{\ast}}=(r_{D}-r)^{-\beta_{D}}(r_{S}-r)^{\beta_{S}}(r+r_{D}+r_{S})^{\beta_{M}}
\ee
where the $\beta$-parameters are as in (\ref{beta}).
The radial null coordinates are\ $U=t-r^{\ast} ,\ V=t+r^{\ast}$\,.
Then the Kruskal coordinates adapted to the interior are
\bea\label{D.75}
& U_{S} = & -e^{-\alpha_{S}(t-r^{\ast})} =
-e^{-\alpha_{S}t} (r_{D}-r)^{-\alpha_{S}\beta_{D}} (r_{S}-r)^{\alpha_{S}\beta_{S}}
(r+r_{D}+r_{S})^{\alpha_{S}\beta_{M}} \ , \nn\\
& V_{S} = & e^{\alpha_{S}(t+r^{\ast})} = e^{\alpha_{S}t} (r_{D}-r)^{-\alpha_{S}\beta_{D}}
(r_{S}-r)^{\alpha_{S}\beta_{S}} (r+r_{D}+r_{S})^{\alpha_{S}\beta_{M}} \ ,
\eea
with $\alpha_{S}=\frac{1}{2\beta_{S}}$.
The reduced two dimensional Schwarzschild de Sitter metric is given by
\be\label{D.76}
ds^{2} = - \lambda\,r \,\frac{dU_{S}dV_{S}}{W^2}\ ,\qquad
W\;=\;\sqrt{r}l\alpha_{S}(r_{D}-r)^{\frac{-(1+2\alpha_{S}\beta_{D})}{2}}(r_{S}-r)^{\frac{2\alpha_{S}\beta_{S}-1}{2}}(r+r_{S}+r_{D})^{\frac{2\alpha_{S}\beta_{M}-1}{2}}\ .
\ee
with $W$ the interior conformal factor.

Towards approximating early times, we will set $t_{a},t_{b} \sim 0$: then 
assuming $b-r_S\gg r_S-a$ as stated above and calculating as earlier
reveals the total entanglement entropy to be
\begin{align}\label{S early}
    S_{total} & =\frac{2\pi a^{2}}{G_{_N}}
    +\frac{c}{6}\log\Big[\frac{2^8r_{S}^4}{(\frac{r_{D}-r_{S}}{l})^4(\frac{2r_{S}+r_{D}}{l})^4}(r_{S}-a)(b-r_{S})(\frac{r_{D}-a}{l})(\frac{r_{D}-b}{l})\cdot \nn\\
    & ~~\qquad\qquad\qquad\qquad\qquad\qquad\qquad\qquad
    (\frac{a+r_{S}+r_{D}}{l})(\frac{b+r_{S}+r_{D}}{l})\Big]\nn\\
    &\qquad +\,\frac{c}{3} \log\Big[1-2\frac{(r_{S}-a)^{\alpha_{S}\beta_{S}}}
        {(b-r_{S})^{\alpha_{S}\beta_{S}}}\,C(a)\Big]\, 
     -\,\frac{c}{3} \log\Big[1+2\frac{(r_{S}-a)^{\alpha_{S}\beta_{S}}}
      {(b-r_{S})^{\alpha_{S}\beta_{S}}}\,C(a) \Big]\ .
\end{align}
The bulk matter entropy is essentially
(\ref{Stotal=area+log.a+a-b+b-.a+b+a-b-/a+b-a-b+}) 
with $a-r_S \ra r_S-a$ as arises using the interior Kruskal variables in
all the distances, and approximating $t_a,\, t_b\sim 0$.\\
Extremizing (\ref{S early}) with respect to the island boundary $a$
gives
\begin{align}\label{extremize 2 }
& \frac{4 \pi a}{G_{_N}}+\frac{c}{6}\Big[-\frac{1}{r_{S}-a}-\frac{1}{r_{D}-a}+\frac{1}{a+r_{S}+r_{D}}\Big]\hfill \nn\\
&\ \ 
  =\, \frac{4c}{6}\frac{\sqrt{\frac{r_{S}-a}{b-r_{S}}}\,C(a)}{1-4\frac{(r_{S}-a)}{(b-r_{S})}\,C(a)^2} \cdot \Big[-\frac{1/2}{r_{S}-a}+\frac{\alpha_{S}\beta_{M}}{a+r_{S}+r_{D}}+\frac{\alpha_{S}\beta_{D}}{r_{D}-a}\Big]\, .
\end{align}


We see that there exists a quantum extremal surface with $0<a\ll r_S$\
and low generalized entropy: approximating (\ref{extremize 2 }) with
$a\ll r_S$ in all $O(c)$ terms gives
\begin{align}
\!\! \frac{4 \pi a}{G_{_N}}  \,\sim\,
     {c\over 6}\,\Big({1\over r_S} + {1\over r_D} - {1\over r_S+r_D}\Big)\,
     +\, {c\over 3}\,\frac{\sqrt{\frac{r_{S}}{b-r_{S}}}\,C(0)}{1-\frac{4r_{S}}{b-r_{S}}\,C(0)^2}\,\Big(-\frac{1}{r_{S}}+\frac{2\alpha_{S}\beta_{M}}{r_{S}+r_{D}}+\frac{2\alpha_{S}\beta_{D}}{r_{D}}\Big)\,.
\end{align}
With $b\gg r_S$, the second set of terms on the right is subleading to
the first so we obtain
\be\label{a value 2}
a \,\sim\,
{G_{_N}c\over 24\pi}\,\Big({1\over r_S} + {1\over r_D} - {1\over r_S+r_D}\Big)\ .
\ee
The quantity in brackets is positive revealing a small quantum extremal
surface $a\sim O(G_{_N}c)$ deep in the interior at early times. It is
worth noting that we have made a coarse approximation in setting
$t_a, t_b\sim 0$: doing this more carefully requires retaining $t_a$ and
extremizing, but we expect similar qualitative behaviour at early times.

For this small QES (\ref{a value 2}), the total on-shell entanglement
entropy approximates to
\begin{align}\label{S_on shell early}
  S_{o.s.} & =\frac{c^{2}G_{_N}}{288 \pi}\Big(\frac{1}{r_{S}}+\frac{1}{r_{D}}-\frac{1}{r_{S}+r_{D}}\Big)^{2} \nn\\
 &\qquad +\, \frac{c}{3}\log\Big[\frac{2^{4}r_{S}^{2}\sqrt{r_{S}(b-r_{S})}}{(\frac{r_{D}-r_{S}}{l})^{2}(\frac{2r_{S}+r_{D}}{l})^{2}} \cdot 
   \sqrt{\frac{r_{D}}{l}\frac{(r_{D}-b)}{l}\frac{(r_{S}+r_{D})}{l}\frac{(b+r_{S}+r_{D})}{l}}\Big]\ ,
\end{align}
ignoring terms scaling as $r_D^{-\al_S\beta_D}$. Thus at early times when
Hawking evaporation has not yet kicked in significantly, the generalized
entropy is not significant, in accordance with the approximate purity of
the early time state.


\begin{thebibliography}{} 

{ \renewcommand{\baselinestretch}{1}
\footnotesize{

\bibitem{Hawking:1976ra}
S.~W.~Hawking,
``Breakdown of Predictability in Gravitational Collapse,''
Phys. Rev. D \textbf{14}, 2460-2473 (1976)
doi:10.1103/PhysRevD.14.2460

\bibitem{Hawking:1975vcx}
S.~W.~Hawking,
``Particle Creation by Black Holes,''
Commun. Math. Phys. \textbf{43}, 199-220 (1975)
[erratum: Commun. Math. Phys. \textbf{46}, 206 (1976)]
doi:10.1007/BF02345020

\bibitem{Page:1993wv}
D.~N.~Page,
``Information in black hole radiation,''
Phys. Rev. Lett. \textbf{71}, 3743-3746 (1993)
doi:10.1103/PhysRevLett.71.3743
[arXiv:hep-th/9306083 [hep-th]].

\bibitem{Page:2013dx}
D.~N.~Page,
``Time Dependence of Hawking Radiation Entropy,''
JCAP \textbf{09}, 028 (2013)
doi:10.1088/1475-7516/2013/09/028
[arXiv:1301.4995 [hep-th]].

\bibitem{Mathur:2009hf}
S.~D.~Mathur,
``The Information paradox: A Pedagogical introduction,''
Class. Quant. Grav. \textbf{26}, 224001 (2009)
doi:10.1088/0264-9381/26/22/224001
[arXiv:0909.1038 [hep-th]].

\bibitem{Almheiri:2012rt}
A.~Almheiri, D.~Marolf, J.~Polchinski and J.~Sully,
``Black Holes: Complementarity or Firewalls?,''
JHEP \textbf{02}, 062 (2013)
doi:10.1007/JHEP02(2013)062
[arXiv:1207.3123 [hep-th]].

\bibitem{Penington:2019npb}
G.~Penington,
``Entanglement Wedge Reconstruction and the Information Paradox,''
JHEP \textbf{09}, 002 (2020)
doi:10.1007/JHEP09(2020)002
[arXiv:1905.08255 [hep-th]].

\bibitem{Almheiri:2019psf}
A.~Almheiri, N.~Engelhardt, D.~Marolf and H.~Maxfield,
``The entropy of bulk quantum fields and the entanglement wedge of an evaporating black hole,''
JHEP \textbf{12}, 063 (2019)
doi:10.1007/JHEP12(2019)063
[arXiv:1905.08762 [hep-th]].

\bibitem{Almheiri:2019hni}
A.~Almheiri, R.~Mahajan, J.~Maldacena and Y.~Zhao,
``The Page curve of Hawking radiation from semiclassical geometry,''
JHEP \textbf{03}, 149 (2020)
doi:10.1007/JHEP03(2020)149
[arXiv:1908.10996 [hep-th]].

\bibitem{Penington:2019kki}
G.~Penington, S.~H.~Shenker, D.~Stanford, Z.~Yang,
``Replica wormholes \& the black hole interior,''
[arXiv:1911.11977[hep-th]].

\bibitem{Almheiri:2019qdq}
A.~Almheiri, T.~Hartman, J.~Maldacena, E.~Shaghoulian and A.~Tajdini,
``Replica Wormholes and the Entropy of Hawking Radiation,''
JHEP \textbf{05}, 013 (2020)
[arXiv:1911.12333 [hep-th]].

\bibitem{Faulkner:2013ana} 
  T.~Faulkner, A.~Lewkowycz and J.~Maldacena,
  ``Quantum corrections to holographic entanglement entropy,''
  JHEP {\bf 1311}, 074 (2013)
  doi:10.1007/JHEP11(2013)074
  [arXiv:1307.2892 [hep-th]].

\bibitem{Engelhardt:2014gca} 
  N.~Engelhardt and A.~C.~Wall,
  ``Quantum Extremal Surfaces: Holographic Entanglement Entropy beyond the Classical Regime,''
  JHEP {\bf 1501}, 073 (2015)
  doi:10.1007/JHEP01(2015)073
  [arXiv:1408.3203 [hep-th]].

\bibitem{Ryu:2006bv} 
  S.~Ryu and T.~Takayanagi,
  ``Holographic derivation of entanglement entropy from AdS/CFT,''
  Phys.\ Rev.\ Lett.\  {\bf 96}, 181602 (2006)
  [hep-th/0603001].

\bibitem{Ryu:2006ef} 
  S.~Ryu and T.~Takayanagi,
  ``Aspects of Holographic Entanglement Entropy,''
  JHEP {\bf 0608}, 045 (2006)
  [hep-th/0605073].

\bibitem{HRT} 
V.~E.~Hubeny, M.~Rangamani and T.~Takayanagi,
``A Covariant holographic entanglement entropy proposal,'' 
JHEP {\bf 0707} (2007) 062  [arXiv:0705.0016 [hep-th]].

\bibitem{Rangamani:2016dms} 
  M.~Rangamani and T.~Takayanagi,
  ``Holographic Entanglement Entropy,''
  Lect.\ Notes Phys.\  {\bf 931}, pp.1 (2017)
  [arXiv:1609.01287 [hep-th]].

\bibitem{Almheiri:2020cfm}
A.~Almheiri, T.~Hartman, J.~Maldacena, E.~Shaghoulian and A.~Tajdini,
``The entropy of Hawking radiation,''
[arXiv:2006.06872 [hep-th]].

\bibitem{Raju:2020smc}
S.~Raju,
``Lessons from the Information Paradox,''
[arXiv:2012.05770 [hep-th]].

\bibitem{Chen:2021lnq}
B.~Chen, B.~Czech and Z.~z.~Wang,
``Quantum Information in Holographic Duality,''
[arXiv:2108.09188 [hep-th]].

\bibitem{Almheiri:2019yqk}
A.~Almheiri, R.~Mahajan and J.~Maldacena,
``Islands outside the horizon,''
[arXiv:1910.11077 [hep-th]].

\bibitem{Chen:2019uhq}
H.~Z.~Chen, Z.~Fisher, J.~Hernandez, R.~C.~Myers and S.~M.~Ruan,
``Information Flow in Black Hole Evaporation,''
JHEP \textbf{03}, 152 (2020)
doi:10.1007/JHEP03(2020)152
[arXiv:1911.03402 [hep-th]].

\bibitem{Almheiri:2019psy}
A.~Almheiri, R.~Mahajan and J.~E.~Santos,
``Entanglement islands in higher dimensions,''
SciPost Phys. \textbf{9}, no.1, 001 (2020)
doi:10.21468/SciPostPhys.9.1.001
[arXiv:1911.09666 [hep-th]].

\bibitem{Gautason:2020tmk}
F.~F.~Gautason, L.~Schneiderbauer, W.~Sybesma and L.~Thorlacius,
``Page Curve for an Evaporating Black Hole,''
JHEP \textbf{05}, 091 (2020)
doi:10.1007/JHEP05(2020)091
[arXiv:2004.00598 [hep-th]].

\bibitem{Anegawa:2020ezn}
T.~Anegawa and N.~Iizuka,
``Notes on islands in asymptotically flat 2d dilaton black holes,''
JHEP \textbf{07}, 036 (2020)
doi:10.1007/JHEP07(2020)036
[arXiv:2004.01601 [hep-th]].

\bibitem{Hashimoto:2020cas}
K.~Hashimoto, N.~Iizuka and Y.~Matsuo,
``Islands in Schwarzschild black holes,''
JHEP \textbf{06}, 085 (2020)
doi:10.1007/JHEP06(2020)085
[arXiv:2004.05863 [hep-th]].

\bibitem{Hartman:2020swn}
T.~Hartman, E.~Shaghoulian and A.~Strominger,
``Islands in Asymptotically Flat 2D Gravity,''
JHEP \textbf{07}, 022 (2020)
doi:10.1007/JHEP07(2020)022
[arXiv:2004.13857 [hep-th]].

\bibitem{Hollowood:2020cou}
T.~J.~Hollowood and S.~P.~Kumar,
``Islands and Page Curves for Evaporating Black Holes in JT Gravity,''
JHEP \textbf{08}, 094 (2020)
doi:10.1007/JHEP08(2020)094
[arXiv:2004.14944 [hep-th]].

\bibitem{Krishnan:2020oun}
C.~Krishnan, V.~Patil and J.~Pereira,
``Page Curve and the Information Paradox in Flat Space,''
[arXiv:2005.02993 [hep-th]].

\bibitem{Alishahiha:2020qza}
M.~Alishahiha, A.~Faraji Astaneh and A.~Naseh,
``Island in the presence of higher derivative terms,''
JHEP \textbf{02}, 035 (2021)
doi:10.1007/JHEP02(2021)035
[arXiv:2005.08715 [hep-th]].

\bibitem{Li:2020ceg}
T.~Li, J.~Chu and Y.~Zhou,
``Reflected Entropy for an Evaporating Black Hole,''
JHEP \textbf{11}, 155 (2020)
doi:10.1007/JHEP11(2020)155
[arXiv:2006.10846 [hep-th]].

\bibitem{Dong:2020uxp}
X.~Dong, X.~L.~Qi, Z.~Shangnan and Z.~Yang,
``Effective entropy of quantum fields coupled with gravity,''
JHEP \textbf{10}, 052 (2020)
doi:10.1007/JHEP10(2020)052
[arXiv:2007.02987 [hep-th]].

\bibitem{Chen:2020jvn}
H.~Z.~Chen, Z.~Fisher, J.~Hernandez, R.~C.~Myers and S.~M.~Ruan,
``Evaporating Black Holes Coupled to a Thermal Bath,''
JHEP \textbf{01}, 065 (2021)
doi:10.1007/JHEP01(2021)065
[arXiv:2007.11658 [hep-th]].

\bibitem{Ling:2020laa}
Y.~Ling, Y.~Liu and Z.~Y.~Xian,
``Island in Charged Black Holes,''
JHEP \textbf{03}, 251 (2021)
doi:10.1007/JHEP03(2021)251
[arXiv:2010.00037 [hep-th]].

\bibitem{Matsuo:2020ypv}
Y.~Matsuo,
``Islands and stretched horizon,''
JHEP \textbf{07}, 051 (2021)
doi:10.1007/JHEP07(2021)051
[arXiv:2011.08814 [hep-th]].

\bibitem{Goto:2020wnk}
K.~Goto, T.~Hartman and A.~Tajdini,
``Replica wormholes for an evaporating 2D black hole,''
JHEP \textbf{04}, 289 (2021)
doi:10.1007/JHEP04(2021)289
[arXiv:2011.09043 [hep-th]].

\bibitem{Akal:2020twv}
I.~Akal, Y.~Kusuki, N.~Shiba, T.~Takayanagi and Z.~Wei,
``Entanglement Entropy in a Holographic Moving Mirror and the Page Curve,''
Phys. Rev. Lett. \textbf{126}, no.6, 061604 (2021)
doi:10.1103/PhysRevLett.126.061604
[arXiv:2011.12005 [hep-th]].

\bibitem{Deng:2020ent}
F.~Deng, J.~Chu and Y.~Zhou,
``Defect extremal surface as the holographic counterpart of Island formula,''
JHEP \textbf{03}, 008 (2021)
doi:10.1007/JHEP03(2021)008
[arXiv:2012.07612 [hep-th]].

\bibitem{Karananas:2020fwx}
G.~K.~Karananas, A.~Kehagias and J.~Taskas,
``Islands in linear dilaton black holes,''
JHEP \textbf{03}, 253 (2021)
doi:10.1007/JHEP03(2021)253
[arXiv:2101.00024 [hep-th]].

\bibitem{Wang:2021woy}
X.~Wang, R.~Li and J.~Wang,
``Islands and Page curves of Reissner-Nordstr\"om black holes,''
JHEP \textbf{04}, 103 (2021)
doi:10.1007/JHEP04(2021)103
[arXiv:2101.06867 [hep-th]].

\bibitem{Verheijden:2021yrb}
E.~Verheijden and E.~Verlinde,
``From the BTZ black hole to JT gravity: geometrizing the island,''
JHEP \textbf{11}, 092 (2021)
doi:10.1007/JHEP11(2021)092
[arXiv:2102.00922 [hep-th]].

\bibitem{Kawabata:2021hac}
K.~Kawabata, T.~Nishioka, Y.~Okuyama and K.~Watanabe,
``Probing Hawking radiation through capacity of entanglement,''
JHEP \textbf{05}, 062 (2021)
doi:10.1007/JHEP05(2021)062
[arXiv:2102.02425 [hep-th]].

\bibitem{Anderson:2020vwi}
L.~Anderson, O.~Parrikar and R.~M.~Soni,
``Islands with gravitating baths: towards ER = EPR,''
JHEP \textbf{21}, 226 (2020)
doi:10.1007/JHEP10(2021)226
[arXiv:2103.14746 [hep-th]].

\bibitem{Bhattacharya:2021jrn}
A.~Bhattacharya, A.~Bhattacharyya, P.~Nandy and A.~K.~Patra,
``Islands and complexity of eternal black hole and radiation subsystems for a doubly holographic model,''
JHEP \textbf{05}, 135 (2021)
doi:10.1007/JHEP05(2021)135
[arXiv:2103.15852 [hep-th]].

\bibitem{Kim:2021gzd}
W.~Kim and M.~Nam,
``Entanglement entropy of asymptotically flat non-extremal and extremal black holes with an island,''
Eur. Phys. J. C \textbf{81}, no.10, 869 (2021)
doi:10.1140/epjc/s10052-021-09680-x
[arXiv:2103.16163 [hep-th]].

\bibitem{Ghosh:2021axl}
K.~Ghosh and C.~Krishnan,
``Dirichlet baths and the not-so-fine-grained Page curve,''
JHEP \textbf{08}, 119 (2021)
doi:10.1007/JHEP08(2021)119
[arXiv:2103.17253 [hep-th]].

\bibitem{Wang:2021mqq}
X.~Wang, R.~Li and J.~Wang,
``Page curves for a family of exactly solvable evaporating black holes,''
Phys. Rev. D \textbf{103}, no.12, 126026 (2021)
doi:10.1103/PhysRevD.103.126026
[arXiv:2104.00224 [hep-th]].

\bibitem{Li:2021lfo}
R.~Li, X.~Wang and J.~Wang,
``Island may not save the information paradox of Liouville black holes,''
Phys. Rev. D \textbf{104}, no.10, 106015 (2021)
doi:10.1103/PhysRevD.104.106015
[arXiv:2105.03271 [hep-th]].

\bibitem{Li:2021mjp}
R.~Li and J.~Wang,
``Hawking radiation and page curves of the black holes in thermal environment,''
Commun. Theor. Phys. \textbf{73}, no.7, 075401 (2021)
doi:10.1088/1572-9494/abf823

\bibitem{Kawabata:2021vyo}
K.~Kawabata, T.~Nishioka, Y.~Okuyama and K.~Watanabe,
``Replica wormholes and capacity of entanglement,''
JHEP \textbf{10}, 227 (2021)
doi:10.1007/JHEP10(2021)227
[arXiv:2105.08396 [hep-th]].

\bibitem{Lu:2021gmv}
Y.~Lu and J.~Lin,
``Islands in Kaluza\textendash{}Klein black holes,''
Eur. Phys. J. C \textbf{82}, no.2, 132 (2022)
doi:10.1140/epjc/s10052-022-10074-w
[arXiv:2106.07845 [hep-th]].

\bibitem{Kruthoff:2021vgv}
J.~Kruthoff, R.~Mahajan and C.~Murdia,
``Free fermion entanglement with a semitransparent interface: the effect of graybody factors on entanglement islands,''
SciPost Phys. \textbf{11}, 063 (2021)
doi:10.21468/SciPostPhys.11.3.063
[arXiv:2106.10287 [hep-th]].

\bibitem{Yu:2021cgi}
M.~H.~Yu and X.~H.~Ge,
``Islands and Page curves in charged dilaton black holes,''
Eur. Phys. J. C \textbf{82}, no.1, 14 (2022)
doi:10.1140/epjc/s10052-021-09932-w
[arXiv:2107.03031 [hep-th]].

\bibitem{Ahn:2021chg}
B.~Ahn, S.~E.~Bak, H.~S.~Jeong, K.~Y.~Kim and Y.~W.~Sun,
``Islands in charged linear dilaton black holes,''
Phys. Rev. D \textbf{105}, no.4, 046012 (2022)
doi:10.1103/PhysRevD.105.046012
[arXiv:2107.07444 [hep-th]].

\bibitem{Wang:2021afl}
X.~Wang, K.~Zhang and J.~Wang,
``What can we learn about islands and state paradox from quantum information theory?,''
[arXiv:2107.09228 [hep-th]].

\bibitem{Cao:2021ujs}
N.~H.~Cao,
``Entanglement entropy and Page curve of black holes with island in massive gravity,''
Eur. Phys. J. C \textbf{82}, no.4, 381 (2022)
doi:10.1140/epjc/s10052-022-10343-8
[arXiv:2108.10144 [hep-th]].

\bibitem{Arefeva:2021kfx}
I.~Aref'eva and I.~Volovich,
``A Note on Islands in Schwarzschild Black Holes,''
[arXiv:2110.04233 [hep-th]].

\bibitem{He:2021mst}
S.~He, Y.~Sun, L.~Zhao and Y.~X.~Zhang,
``The universality of islands outside the horizon,''
JHEP \textbf{05}, 047 (2022)
doi:10.1007/JHEP05(2022)047
[arXiv:2110.07598 [hep-th]].

\bibitem{Matsuo:2021mmi}
Y.~Matsuo,
``Entanglement entropy and vacuum states in Schwarzschild geometry,''
JHEP \textbf{06}, 109 (2022)
doi:10.1007/JHEP06(2022)109
[arXiv:2110.13898 [hep-th]].

\bibitem{Tian:2022pso}
J.~Tian,
``Islands in Generalized Dilaton Theories,''
[arXiv:2204.08751 [hep-th]].

\bibitem{Laddha:2020kvp}
A.~Laddha, S.~G.~Prabhu, S.~Raju and P.~Shrivastava,
``The Holographic Nature of Null Infinity,''
SciPost Phys. \textbf{10}, no.2, 041 (2021)
doi:10.21468/SciPostPhys.10.2.041
[arXiv:2002.02448 [hep-th]].

\bibitem{Geng:2021hlu}
H.~Geng, A.~Karch, C.~Perez-Pardavila, S.~Raju, L.~Randall, M.~Riojas and S.~Shashi,
``Inconsistency of Islands in Theories with Long-Range Gravity,''
[arXiv:2107.03390 [hep-th]].

\bibitem{Bena:2022rna}
I.~Bena, E.~J.~Martinec, S.~D.~Mathur and N.~P.~Warner,
``Fuzzballs and Microstate Geometries: Black-Hole Structure in String Theory,''
[arXiv:2204.13113 [hep-th]].

\bibitem{Gibbons:1977mu} 
  G.~W.~Gibbons and S.~W.~Hawking,
  ``Cosmological Event Horizons, Thermodynamics, and Particle Creation,''
  Phys.\ Rev.\ D {\bf 15}, 2738 (1977).
  doi:10.1103/PhysRevD.15.2738

\bibitem{Ginsparg:1982rs} 
  P.~H.~Ginsparg and M.~J.~Perry,
  ``Semiclassical Perdurance of de Sitter Space,''
  Nucl.\ Phys.\ B {\bf 222}, 245 (1983).
  doi:10.1016/0550-3213(83)90636-3

\bibitem{Bousso:1996au} 
  R.~Bousso and S.~W.~Hawking,
  ``Pair creation of black holes during inflation,''
  Phys.\ Rev.\ D {\bf 54}, 6312 (1996)
  doi:10.1103/PhysRevD.54.6312
  [gr-qc/9606052].

\bibitem{Bousso:1997wi}
R.~Bousso and S.~W.~Hawking,
``(Anti)evaporation of Schwarzschild-de Sitter black holes,''
Phys. Rev. D \textbf{57}, 2436-2442 (1998)
doi:10.1103/PhysRevD.57.2436
[arXiv:hep-th/9709224 [hep-th]].

\bibitem{Nariai}
  H. Nariai,
  ``On some static solutions of Einstein’s gravitational field equations
  in a spherically symmetric case'',
  Sci. Rep. Tohoku Univ. Eighth Ser. 34, 1950.

\bibitem{Maldacena:2019cbz} 
  J.~Maldacena, G.~J.~Turiaci and Z.~Yang,
  ``Two dimensional Nearly de Sitter gravity,''
  arXiv:1904.01911 [hep-th].

\bibitem{Anninos:2012ft} 
  D.~Anninos, F.~Denef and D.~Harlow,
  ``The Wave Function of Vasiliev's Universe - A Few Slices Thereof,''
  Phys.\ Rev.\ D {\bf 88}, 084049 (2013)
  [arXiv:1207.5517 [hep-th]].

\bibitem{Fernandes:2019ige}
K.~Fernandes, K.~S.~Kolekar, K.~Narayan and S.~Roy,
``Schwarzschild de Sitter and extremal surfaces,''
Eur. Phys. J. C \textbf{80}, no.9, 866 (2020)
doi:10.1140/epjc/s10052-020-08437-2
[arXiv:1910.11788 [hep-th]].

\bibitem{Shankaranarayanan:2003ya}
S.~Shankaranarayanan,
``Temperature and entropy of Schwarzschild-de Sitter space-time,''
Phys. Rev. D \textbf{67}, 084026 (2003)
doi:10.1103/PhysRevD.67.084026
[arXiv:gr-qc/0301090 [gr-qc]].

\bibitem{Guven:1990ubi}
J.~Guven and D.~N\'u\~nez,
``Schwarzschild-de Sitter space and its perturbations,''
Phys. Rev. D \textbf{42}, no.8, 2577-2584 (1990)
doi:10.1103/physrevd.42.2577

\bibitem{Strominger:1994tn}
A.~Strominger,
``Les Houches lectures on black holes,''
[arXiv:hep-th/9501071 [hep-th]].

\bibitem{Grumiller:2002nm}
D.~Grumiller, W.~Kummer and D.~V.~Vassilevich,
``Dilaton gravity in two-dimensions,''
Phys. Rept. \textbf{369}, 327-430 (2002)
doi:10.1016/S0370-1573(02)00267-3
[arXiv:hep-th/0204253 [hep-th]].

\bibitem{Narayan:2020pyj}
K.~Narayan,
``On aspects of two-dimensional dilaton gravity, dimensional reduction, and holography,''
Phys. Rev. D \textbf{104}, no.2, 026007 (2021)
doi:10.1103/PhysRevD.104.026007
[arXiv:2010.12955 [hep-th]].

\bibitem{Bhattacharya:2020qil}
R.~Bhattacharya, K.~Narayan and P.~Paul,
``Cosmological singularities and 2-dimensional dilaton gravity,''
JHEP \textbf{08}, 062 (2020)
doi:10.1007/JHEP08(2020)062
[arXiv:2006.09470 [hep-th]].

\bibitem{Calabrese:2004eu}
P.~Calabrese and J.~L.~Cardy,
``Entanglement entropy and quantum field theory,''
J. Stat. Mech. \textbf{0406}, P06002 (2004)
doi:10.1088/1742-5468/2004/06/P06002
[arXiv:hep-th/0405152 [hep-th]].

\bibitem{Calabrese:2009qy}
P.~Calabrese and J.~Cardy,
``Entanglement entropy and conformal field theory,''
J. Phys. A \textbf{42}, 504005 (2009)
doi:10.1088/1751-8113/42/50/504005
[arXiv:0905.4013 [cond-mat.stat-mech]].

\bibitem{Calabrese:2009ez}
P.~Calabrese, J.~Cardy, E.~Tonni,
``Entanglement entropy of two disjoint intervals in conformal field theory,''
J. Stat. Mech. \textbf{0911}, P11001 (2009)\! 
doi:10.1088/1742-5468/2009/11/P11001\!
[arXiv:0905.2069\,[hep-th]].

\bibitem{Calabrese:2010he}
P.~Calabrese, J.~Cardy, E.~Tonni,
``Entanglement entropy of two disjoint intervals in conformal field theory II,''
J. Stat. Mech. \textbf{1101}, P01021 (2011)\!
doi:10.1088/1742-5468/2011/01/P01021\!
[arXiv:1011.5482\,[hep-th]].

\bibitem{Headrick:2010zt} 
  M.~Headrick,
  ``Entanglement Renyi entropies in holographic theories,''
  Phys.\ Rev.\ D {\bf 82}, 126010 (2010)
  [arXiv:1006.0047 [hep-th]].

\bibitem{Ageev:2022hqc}
D.~S.~Ageev and I.~Y.~Aref'eva,
``Thermal density matrix breaks down the Page curve,''
[arXiv:2206.04094 [hep-th]].

\bibitem{Krishnan:2020fer}
C.~Krishnan,
``Critical Islands,''
JHEP \textbf{01}, 179 (2021)
doi:10.1007/JHEP01(2021)179
[arXiv:2007.06551 [hep-th]].

\bibitem{Chen:2020tes}
Y.~Chen, V.~Gorbenko and J.~Maldacena,
``Bra-ket wormholes in gravitationally prepared states,''
[arXiv:2007.16091 [hep-th]].

\bibitem{Hartman:2020khs}
T.~Hartman, Y.~Jiang and E.~Shaghoulian,
``Islands in cosmology,''
JHEP \textbf{11}, 111 (2020)
doi:10.1007/JHEP11(2020)111
[arXiv:2008.01022 [hep-th]].

\bibitem{VanRaamsdonk:2020tlr}
M.~Van Raamsdonk,
``Comments on wormholes, ensembles, and cosmology,''
arXiv:2008.02259[hep-th].

\bibitem{Balasubramanian:2020xqf}
V.~Balasubramanian, A.~Kar and T.~Ugajin,
``Islands in de Sitter space,''
JHEP \textbf{02}, 072 (2021)
doi:10.1007/JHEP02(2021)072
[arXiv:2008.05275 [hep-th]].

\bibitem{Sybesma:2020fxg}
W.~Sybesma,
``Pure de Sitter space and the island moving back in time,''
Class. Quant. Grav. \textbf{38}, no.14, 145012 (2021)
doi:10.1088/1361-6382/abff9a
[arXiv:2008.07994 [hep-th]].

\bibitem{Manu:2020tty}
A.~Manu, K.~Narayan and P.~Paul,
``Cosmological singularities, entanglement and quantum extremal surfaces,''
JHEP \textbf{04}, 200 (2021)
doi:10.1007/JHEP04(2021)200
[arXiv:2012.07351 [hep-th]].

\bibitem{Choudhury:2020hil}
S.~Choudhury, S.~Chowdhury, N.~Gupta, A.~Mishara, S.~P.~Selvam, S.~Panda, G.~D.~Pasquino, C.~Singha and A.~Swain,
``Circuit Complexity From Cosmological Islands,''
Symmetry \textbf{13}, 1301 (2021)
doi:10.3390/sym13071301
[arXiv:2012.10234 [hep-th]].

\bibitem{Bousso:2021sji}
R.~Bousso and A.~Shahbazi-Moghaddam,
``Island Finder and Entropy Bound,''
Phys. Rev. D \textbf{103}, no.10, 106005 (2021)
doi:10.1103/PhysRevD.103.106005
[arXiv:2101.11648 [hep-th]].

\bibitem{Geng:2021wcq}
H.~Geng, Y.~Nomura and H.~Y.~Sun,
``Information paradox and its resolution in de Sitter holography,''
Phys. Rev. D \textbf{103}, no.12, 126004 (2021)
doi:10.1103/PhysRevD.103.126004
[arXiv:2103.07477 [hep-th]].

\bibitem{Fallows:2021sge}
S.~Fallows and S.~F.~Ross,
``Islands and mixed states in closed universes,''
JHEP \textbf{07}, 022 (2021)
doi:10.1007/JHEP07(2021)022
[arXiv:2103.14364 [hep-th]].

\bibitem{Aalsma:2021bit}
L.~Aalsma and W.~Sybesma,
``The Price of Curiosity: Information Recovery in de Sitter Space,''
JHEP \textbf{05}, 291 (2021)
doi:10.1007/JHEP05(2021)291
[arXiv:2104.00006 [hep-th]].

\bibitem{Giataganas:2021cwg}
D.~Giataganas and N.~Tetradis,
``Entanglement entropy in FRW backgrounds,''
Phys. Lett. B \textbf{820}, 136493 (2021)
doi:10.1016/j.physletb.2021.136493
[arXiv:2105.12614 [hep-th]].

\bibitem{Aalsma:2021kle}
L.~Aalsma, A.~Cole, E.~Morvan, J.~P.~van der Schaar and G.~Shiu,
``Shocks and information exchange in de Sitter space,''
JHEP \textbf{10}, 104 (2021)
doi:10.1007/JHEP10(2021)104
[arXiv:2105.12737 [hep-th]].

\bibitem{Langhoff:2021uct}
K.~Langhoff, C.~Murdia and Y.~Nomura,
``Multiverse in an inverted island,''
Phys. Rev. D \textbf{104}, no.8, 086007 (2021)
doi:10.1103/PhysRevD.104.086007
[arXiv:2106.05271 [hep-th]].

\bibitem{Aguilar-Gutierrez:2021bns}
S.~E.~Aguilar-Gutierrez, A.~Chatwin-Davies, T.~Hertog, N.~Pinzani-Fokeeva and B.~Robinson,
``Islands in Multiverse Models,''
[arXiv:2108.01278 [hep-th]].

\bibitem{Shaghoulian:2021cef}
E.~Shaghoulian,
``The central dogma and cosmological horizons,''
JHEP \textbf{01}, 132 (2022)
doi:10.1007/JHEP01(2022)132
[arXiv:2110.13210 [hep-th]].

\bibitem{Goswami:2021ksw}
K.~Goswami, K.~Narayan and H.~K.~Saini,
``Cosmologies, singularities and quantum extremal surfaces,''
JHEP \textbf{03}, 201 (2022)
doi:10.1007/JHEP03(2022)201
[arXiv:2111.14906 [hep-th]].

\bibitem{Bousso:2022gth}
R.~Bousso and E.~Wildenhain,
``Islands in closed and open universes,''
Phys. Rev. D \textbf{105}, no.8, 086012 (2022)
doi:10.1103/PhysRevD.105.086012
[arXiv:2202.05278 [hep-th]].

\bibitem{Narayan:2020nsc}
K.~Narayan,
``de Sitter future-past extremal surfaces and the entanglement wedge,''
Phys. Rev. D \textbf{101}, no.8, 086014 (2020)
doi:10.1103/PhysRevD.101.086014
[arXiv:2002.11950 [hep-th]].

\bibitem{Kames-King:2021etp}
J.~Kames-King, E.~M.~H.~Verheijden and E.~P.~Verlinde,
``No Page curves for the de Sitter horizon,''
JHEP \textbf{03}, 040 (2022)
doi:10.1007/JHEP03(2022)040
[arXiv:2108.09318 [hep-th]].

\bibitem{Moitra:2022glw}
U.~Moitra, S.~K.~Sake and S.~P.~Trivedi,
``Aspects of Jackiw-Teitelboim gravity in Anti-de Sitter and de Sitter spacetime,''
JHEP \textbf{06}, 138 (2022)
doi:10.1007/JHEP06(2022)138
[arXiv:2202.03130 [hep-th]].

\bibitem{Svesko:2022txo}
A.~Svesko, E.~Verheijden, E.~P.~Verlinde and M.~R.~Visser,
``Quasi-local energy and microcanonical entropy in two-dimensional nearly de Sitter gravity,''
[arXiv:2203.00700 [hep-th]].

\bibitem{Levine:2022wos}
A.~Levine and E.~Shaghoulian,
``Encoding beyond cosmological horizons in de Sitter JT gravity,''
[arXiv:2204.08503 [hep-th]].

\bibitem{Chandrasekaran:2022cip}
V.~Chandrasekaran, R.~Longo, G.~Penington and E.~Witten,
``An Algebra of Observables for de Sitter Space,''
[arXiv:2206.10780 [hep-th]].

} }
\end{thebibliography}
\end{document}